\documentclass[%
 aip,
jap,%
 amsmath,amssymb,
 reprint,%
]{revtex4-1}

\usepackage{hyperref}
\hypersetup{colorlinks=true, allcolors=blue}

\usepackage{physics}
\usepackage{graphicx}
\usepackage{dcolumn}
\usepackage{bm}
\usepackage{textcomp}
\usepackage{amsmath}
\usepackage{mathtools}

\def\mathclap#1{\text{\hbox to 0pt{\hss$\mathsurround=0pt#1$\hss}}}

\newcommand*{\citen}[1]{%
  \begingroup
    \romannumeral-`\x 
    \setcitestyle{numbers}%
    \cite{#1}%
  \endgroup   
}

\makeatletter
\newcommand{\vast}{\bBigg@{5}}
\newcommand{\Vast}{\bBigg@{5}}
\makeatother

\usepackage{xcolor}

\newcommand*{\colorboxed}{}
\def\colorboxed#1#{%
  \colorboxedAux{#1}%
}

\newcommand*{\colorboxedAux}[3]{%
  \begingroup
    \colorlet{cb@saved}{.}%
    \color#1{#2}%
    \boxed{%
      \color{cb@saved}%
      #3%
    }%
  \endgroup
}

\usepackage{subfigure}

\usepackage{soul}

\usepackage{multirow}

\begin{document}

\preprint{AIP/123-QED}

\title[]{Switching time of spin-torque-driven magnetization in biaxial ferromagnets}

\author{Ankit Shukla}
\email{ankits4@illinois.edu}
\affiliation{ 
Holonyak Micro and Nanotechnology Laboratory, University of Illinois at Urbana-Champaign, Urbana, IL 61801
}
\author{Arun Parthasarathy}
\email{arun.parth@nyu.edu}
\affiliation{ 
Department of Electrical and Computer Engineering, New York University, Brooklyn 11201, USA
}
\author{Shaloo Rakheja}
\email{rakheja@illinois.edu}
\affiliation{ 
Holonyak Micro and Nanotechnology Laboratory, University of Illinois at Urbana-Champaign, Urbana, IL 61801
}

\date{\today}

\begin{abstract}
We analytically model the magnetization switching time of a biaxial ferromagnet driven by an antidamping-like spin torque. The macrospin magnetization dynamics is mapped to an energy-flow equation, wherein a rational-function approximation of the elliptic integrals for moderate spin current and small damping results in a closed-form expression of the switching time. Randomness in the initial angle of magnetization gives the distribution function of the switching time. The analytic model conforms to the results obtained from Monte Carlo simulation for a broad range of material parameters. Our results can ameliorate design and benchmarking of in-plane spin torque magnetic memory by obviating expensive numerical computation.
\end{abstract}

\maketitle

\section{Introduction}
\begin{figure}[ht!]
    \centering{ \label{3}
    \includegraphics[width=\columnwidth, clip = true,trim = 0mm 0mm 0mm 0mm]{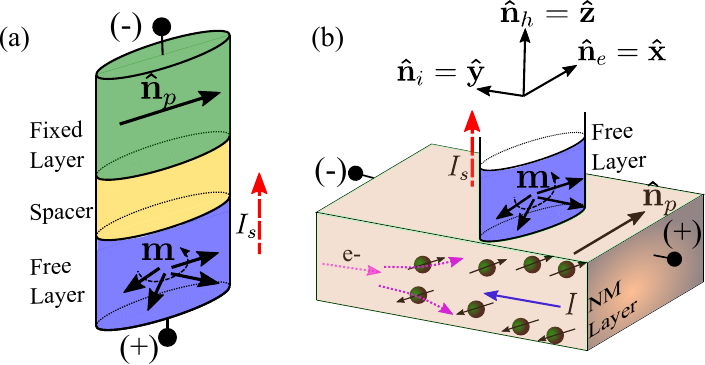}}
    \caption{Injection of spin current $I_{s}$ (dashed red line) into a free magnetic layer produces (a) STT in a spin valve and (b) SOT in a spin-Hall structure. $\mathbf{m}$ represents the direction of free-layer magnetization, $\mathbf{\hat{n}}_{p}$ represents the direction of spin polarization, and $\mathbf{\hat{n}}_{e}$, $\mathbf{\hat{n}}_{i}$, and $\mathbf{\hat{n}}_{h}$ are the unit vectors along the easy, intermediate, and hard anisotropy axes, respectively. The flow of electron in the nonmagnetic (NM) layer is opposite to that of the injected charge current $I$.}
    \label{memory}
\end{figure}

Current-induced spin phenomena, such as spin-transfer torque (STT) and spin-orbit torque (SOT), allow for electrical manipulation of the magnetic order, and form the basis of emerging spintronic technologies such has non-volatile memory,~\cite{kent2015new} magnonic interconnects,~\cite{sadovnikov2017toward} and radio-frequency oscillators.~\cite{chen2016spin}
Spin current can transfer angular momentum to a magnetic layer and reorient its magnetization, similar to how electric current can transfer charge to a capacitor and modulate its voltage. 
As shown in Fig.~\ref{memory}(a), an electric current flowing orthogonal to the plane of the spin valve becomes spin polarized in a direction parallel to the magnetization of the fixed layer. This spin-polarized current affects the magnetization of the free layer due to STT.~\cite{slonczewski1996current, berger1996emission} On the other hand, an in-plane electric current flowing through a nonmagnetic \textcolor{black}{(NM)} material with spin-orbit coupling is spin polarized in the plane of the nonmagnetic material, but transverse to the electric current due to the spin-Hall effect.~\cite{sinova2015spin} This in-plane polarized spin current can exert SOT~\cite{liu2012spin} on the free-layer magnetization of the spin valve as shown in Fig.~\ref{memory}(b). 

Thin-film magnets---the path toward miniaturized spintronics---are subject to epitaxial strain from substrate and finite-size effects, which can elicit an in-plane or a perpendicular spin orientation.~\cite{sander2004magnetic} The symmetries in the energy landscape of thin films up to quadratic (lowest-order) terms of magnetization components are characterized by a biaxial anisotropy, consisting of an axis of minimum energy that is `easy' for spins to orient along and an orthogonal axis of maximum energy which is `hard'. The uniaxial anisotropy is a special case of the biaxial anisotropy where the hard axis is absent. For perpendicularly magnetized films, a perpendicular easy axis approximates the anisotropy by assuming symmetry in the plane. For in-plane magnetized films, an in-plane easy axis and a perpendicular hard axis offer the correct description.

In STT memory, the perpendicular free-layer configuration is superior to the in-plane configuration due to its lower switching current, faster speeds, and higher density.~\cite{kent2015new} A key problem in writing STT memory is its vulnerability to dielectric breakdown of the tunnel barrier. This is addressed in SOT memory, where the writing occurs with an in-plane current that need not traverse the tunnel barrier. The three-terminal SOT memory separates the read and write paths, improving memory endurance at the cost of cell size.
However, deterministic switching of the perpendicular free layer in SOT memory requires either a biasing magnetic field~\cite{liu2012current} or additional layers in the device stack adding to its fabrication complexity.~\cite{oh2016field} The in-plane free-layer configuration of the SOT memory is preferred due to its fabrication simplicity, magnetic-field-free switching, and lower switching currents,~\cite{fukami2016spin} although its writing speed is inferior to that of the perpendicular SOT memory.  

When the spin polarization of the injected spin current is antiparallel to the stable orientation of the free-layer magnetization, the spin torque is antidamping-like---it competes with the intrinsic damping to raise the macrospin energy---until halfway in the magnetization reversal process when it becomes damping-like---it contributes to damping to cause dissipation of macrospin energy. The switching process is characterized by the time to reverse the orientation of the free layer as a function of the input spin current. 
A closed-form expression of the switching time is useful to design and optimize the performance of spin-torque memory. Previous works~\citep{sun2000spin, liu2012spin, pinna2013spin, d2019analysis} have derived expressions of the switching time for the uniaxial anisotropy, but not for the more general biaxial anisotropy presented in this work. 

\textcolor{black}{Here, we treat the magnetization of the free-layer ferromagnet as a single-domain or a macrospin,\citep{brown1963thermal} which switches coherently due to spin torque. The macrospin approximation is valid if the dimensions of the ferromagnet are smaller than the characteristic domain-wall width, typically $\sim$ 50 nm for Co, Fe, Ni.~\citep{park2014macrospin, moreno2016temperature} 
The macrospin model has been applied successfully to explain measured probability distribution of switching dynamics in ferromagnetic thin films with dimensions on the order of 50 nm.\citep{bedau2010spin}   
Inhomogeneous switching dynamics, such as due to domain wall nucleation and propagation, have been experimentally measured in large ferromagnetic samples.~\citep{baumgartner2017spatially, grimaldi2020single} However, consideration of multi-domain switching of the ferromagnet is out of scope of this work.}


\textcolor{black}{Within the macrospin approximation} there are three equivalent approaches to analyze the magnetization dynamics: perturbative approach,~\citep{sun2000spin} constructing a Fokker-Planck representation,~\citep{apalkov2005spin,  taniguchi2013spin} or using a constant-energy stochastic equation.~\citep{mayergoyz2009nonlinear, pinna2013thermally, newhall2013averaged, pinna2014spin} In this work, we adopt the constant-energy orbit averaging (CEOA) to study magnetization reversal in a biaxial magnetic system because this approach simplifies
a coupled three-dimensional (3D) (two-dimensional if the magnetization has a constant magnitude) stochastic problem into a tractable one-dimensional (1D) problem at low temperature and for low-to-moderate applied spin current.~\citep{pinna2013thermally, newhall2013averaged, pinna2014spin}  

The switching dynamics is modeled as a slow perturbation of the rapid constant-energy gyration around the easy axis (Sec.~\ref{sec:theory}). In the deterministic limit, a closed-form expression of the switching time as a function of input spin current, initial magnetization energy, and material parameters is obtained. Average switching time and the probability distribution of the switching time follow for an initial Boltzmann distributed ensemble of spins (Sec.~\ref{sec:analytic}). \textcolor{black}{Analytic results are shown to conform to those obtained from numerical Monte Carlo simulations. A brief discussion of the stochastic dynamics and its applications is included in Sec.~\ref{sec:results_disc}.}

\section{Theory}\label{sec:theory}
The dynamics of magnetization subject to an effective magnetic field, intrinsic damping, and spin torque is described by the Landau-Lifshitz-Gilbert (LLG) equation. Using dimensionless form of physical parameters listed in Table~\ref{tab:symbs}, the LLG equation is~\citep{pinna2013thermally, yu2019importance}
\begin{equation}\label{sLLGS}
    \begin{split}
        \frac{\partial{\mathbf{m}}}{\partial \tau} &= - \left(\mathbf{m} \times \mathbf{h}_\mathrm{eff}\right) - \alpha \mathbf{m} \times \left(\mathbf{m} \times \mathbf{h}_\mathrm{eff}\right) \\  
        & - I_{s}\mathbf{m} \times \left(\mathbf{m} \times \mathbf{\hat{n}}_{p}\right) + \alpha I_{s}\left(\mathbf{m} \times \mathbf{\hat{n}}_{p}\right),
    \end{split}
\end{equation}
where $\mathbf{m}$ is the normalized free-layer magnetization, $\alpha$ is the Gilbert damping constant, $I_{s}$ is the input spin current, and $\mathbf{\hat{n}}_{p}$ is the unit vector along the direction of spin polarization.  
\begin{table}[t!]
\caption{\label{tab:symbs} Symbols and Definitions in Eq.~(\ref{sLLGS}). $M_s$, $H_k$, and $V$ are the saturation magnetization, Stoner-Wohlfarth field, and volume, respectively, of the free layer. $\mu_0$ is the free-space permeability, $\gamma$ is the gyromagnetic ratio, and $\alpha$ is the Gilbert damping constant.}
\begin{ruledtabular}
\begin{tabular}{ll}
Physical parameters  &  Dimensionless form \\
\hline
Magnetization $\mathbf{M}$ & $\mathbf{m} = \mathbf{M}/M_{s}$\\[1pt] 
Time $t$ & $\tau = \gamma \mu_{0} H_{k} t /(1 + \alpha^{2})$\\[1pt] 
Effective magnetic field $\mathbf{H}_\mathrm{eff}$ &  $\mathbf{h}_\mathrm{eff} = \mathbf{H}_\mathrm{eff}/H_{k}$ \\[1pt]
Spin current $\beta_{s}$ & $I_{s} = \beta_{s}/(\mu_{0} M_{s} H_{k} V)$
\end{tabular}
\end{ruledtabular}
\end{table}
The effective magnetic field, $\mathbf{h}_\mathrm{eff}$, includes contributions from an internal field produced by the magnetic anisotropy, externally applied magnetic fields ($\mathbf{H}_{a}$), and a thermal field. 
The sum of the internal and external magnetic fields, 
normalized to the Stoner-Wohlfarth field, $H_k$, is given as 
\begin{equation}\label{eq:h}
     \mathbf{h} = -\frac{1}{2}\nabla_{\mathbf{m}}\left[g_{L}(\mathbf{m}, \mathbf{H}_a)\right],
\end{equation}
where $g_L$ is the free energy of the macrospin normalized to its uniaxial energy $K_{u} V$, where $K_{u} = \mu_{0} M_{s} H_{k}/2$ ($\mu_0$ is the free-space permeability and $M_s$ is the saturation magnetization of the free layer). Neglecting higher order anisotropy terms,
\begin{equation}\label{energy}
    \begin{split}
        g_{L}(\mathbf{m}, \mathbf{H}_{a}) &=  D_{e} \left(\mathbf{m} \cdot \mathbf{\hat{n}}_{e}\right)^{2} + D_{h} \left(\mathbf{m} \cdot \mathbf{\hat{n}}_{h}\right)^{2} - \frac{2}{H_{k}}\mathbf{m} \cdot \mathbf{H}_{a}, 
    \end{split}
\end{equation}
where
$D_{e}$ and $D_{h}$ are the effective anisotropy coefficients along the easy $(\mathbf{\hat{n}}_{e})$ and hard axes $(\mathbf{\hat{n}}_{h})$, respectively. For thin-film magnets, 
the energy landscape is characterized by a biaxial anisotropy with 
$D_{e} = -1$ and $D_{h}= M_{s}/H_{k}$. 
Without loss of generality, $\mathbf{\hat{n}}_{e}$ and $\mathbf{\hat{n}}_{h}$ are assumed to coincide with $\mathbf{\hat{x}}$ and $\mathbf{\hat{z}}$ axes, respectively.  
Using Eq.~(\ref{energy}) in Eq.~(\ref{eq:h}), we obtain $\mathbf{h} = m_{x} \mathbf{\hat{x}} - R m_{z} \mathbf{\hat{z}} + \mathbf{H}_{a}/H_{k}$ where $R = D_{h} = M_{s}/H_{k}$.  

The thermal field is a Langevin field~\citep{kani2017modeling} that is spatially isotropic and uncorrelated in space and time, therefore,
\begin{subequations}\label{thermal}
    \begin{flalign}
        \langle \mathbf{h}_{T}(t)\rangle &= 0, \\
        \langle \mathbf{h}_{T, p}(t_{1}) \mathbf{h}_{T, r}(t_{2}) \rangle &= D \delta_{p, r} \delta(t_{1} - t_{2}),
    \end{flalign}
\end{subequations}
where $p, r$ represent Cartesian coordinates, $\delta_{p, r}$ is the Kronecker delta, and $\delta(t)$ is the Dirac delta. 
Assuming the macrospin to be in thermal equilibrium with the thermal bath and neglecting Joule heating, the diffusion coefficient
\begin{equation}\label{diffusivity}
    D = \frac{\alpha k_{B} T}{\left(1+\alpha^{2}\right)K_{u} V} = \frac{\alpha}{\left(1+\alpha^{2}\right)\Delta_{0}},
\end{equation}
where $k_{B}$ is the Boltzmann constant, $T$ is the temperature of the bath, and \textcolor{black}{$\Delta_{0}$ is the barrier height of a uniaxial anisotropy magnet and measures the thermal stability of the macrospin.~\citep{kani2017modeling} Here, $\Delta_{0}$ is normalized to thermal energy $k_BT$.}

\vspace{-5pt}
\subsection{Constant-Energy Orbit Averaging (CEOA)}
\begin{figure}[ht!]
    \centering
    \includegraphics[width=\columnwidth, clip = true, trim = 0mm 10mm 10mm 35mm]{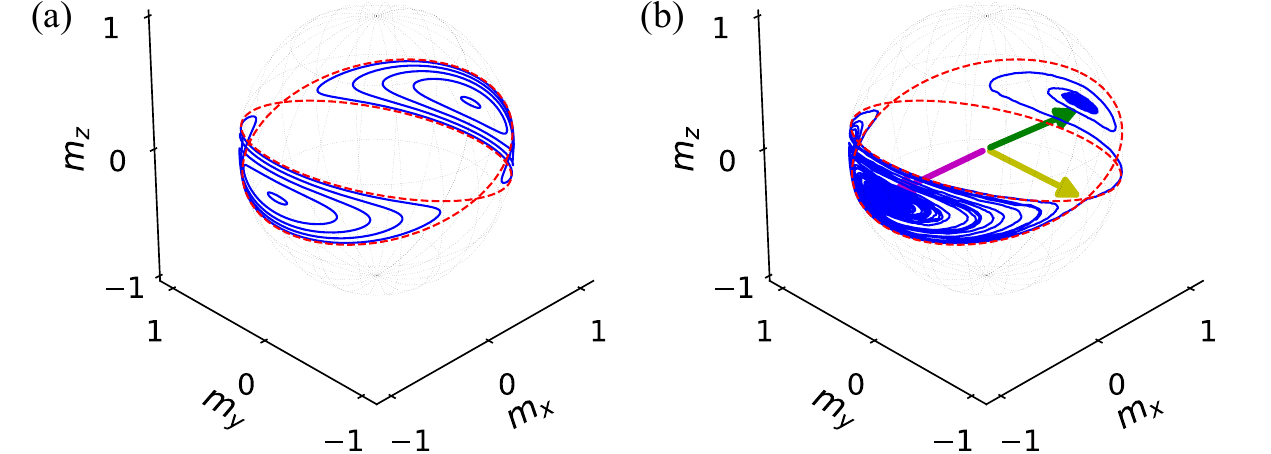}
    \caption{(a) Constant-energy curves for $g_{L} < 0$ in both anti-parallel and parallel wells for $R = 15$. (b) Switching trajectory of a magnetization initially in the anti-parallel well for $R = 15$ and a large input spin current. External field is assumed to be absent. The arrows shown in the figure mark the initial (magenta), intermediate (yellow) and final (green) positions of the magnetization, in the counter-clockwise sense. The dashed red curves in both figures correspond to a zero energy separatrix.}
    \label{fig:const_and_switch}
\end{figure}

\vspace{-10pt}
The energy of a macrospin is conserved when damping, thermal field (noise), external magnetic field, and spin torque are absent. In this case, $\mathbf{m}$ precesses around the easy axis on the unit magnetization sphere with a fixed macrospin energy $g_{L}$ $(< 0)$. Trajectories of conserved motion, illustrated in Fig.~\ref{fig:const_and_switch}(a), are obtained by solving Eqs.~(\ref{sLLGS}) and~(\ref{energy}) with $\alpha$, $\mathbf{H}_a$, and $I_s$ set equal to zero, and $\norm{\mathbf{m}} = 1$. However, with finite damping, $\mathbf{m}$ loses energy, eventually relaxing to a stable equilibrium state ($m_x = \pm 1$). A non-zero spin torque can pump energy to the macrospin and act against its inherent damping, causing the magnetization to deviate from its equilibrium position.

{In the case of a small to moderate input spin torque, small damping, and low temperature, two distinct time-scales of magnetization dynamics emerge: (i) a fast time-scale}
associated with constant-energy gyration around the easy axis and (ii) a slow time-scale corresponding to perpendicular diffusion of magnetization from one constant-energy orbit to another as a result of damping, spin torque, and themal field. Figure~\ref{fig:const_and_switch}(b) shows one such trajectory where the magnetization switches from an anti-parallel well to a parallel well under the simultaneous effects of damping, input spin current, and thermal field 

The rate of change of macrospin energy due to the non-conservative torques is given as
\begin{equation}\label{gdot}
    \frac{ \partial g_{L}}{\partial \tau} = \nabla_{\mathbf{m}} g_{L} \cdot \frac{\partial \mathbf{{m}}}{\partial \tau} = 2\left(R m_{z} \frac{\partial m_{z}}{\partial \tau} - m_{x}\frac{\partial m_{x}}{\partial \tau}\right).
\end{equation}
Averaging the above equation over one time period of the undamped motion reduces the coupled stochastic dynamics of Eq.~(\ref{sLLGS}) to a 1D stochastic dynamics as~\citep{pinna2013thermally, pinna2014spin} (see Appendix~\ref{derive_gdot})
\begin{align}\label{avg}
    \begin{split}
        &\left \langle \frac{\partial g_{L}}{\partial \tau} \right \rangle = \frac{\pi \alpha}{K\left(R, g_{L}\right)} \sqrt{\frac{R-g_{L}}{1+R}} \left[\frac{I_{s}}{\alpha} \left(1+g_{L}\right)\right. \\
        & - \left. \frac{2}{\pi} \sqrt{\left(1+R\right) \left(R-g_{L}\right)}\left\{E\left(R, g_{L}\right) + g_{L} K\left(R, g_{L}\right)\right \}\right] \\
        &+ 2\sqrt{\frac{\alpha}{\Delta_{0}}} \sqrt{\frac{\left(R-g_{L}\right)}{K\left(R, g_{L}\right)}} \sqrt{ E\left(R, g_{L}\right) + g_{L} K\left(R, g_{L}\right)}\circ \dot{W}_{g_{L}}.
    \end{split}
\end{align}
Here, $\circ$ denotes multiplication of thermal noise in the Stratonovich sense,~\citep{pinna2013thermally,kani2017modeling} while $K\left(R, g_{L}\right)$ and $E\left(R, g_{L}\right)$ are the complete elliptic integrals of the first and second kind, respectively. Time averaging $\frac{ \partial g_{L}}{\partial \tau}$ over a period of precessional motion enables us to study the dynamics due to slow diffusion of energy with respect to fast periodic oscillations.~\citep{pinna2013thermally,pinna2014spin}

\begin{figure*}[ht!]
  \centering
  \includegraphics[width = 2\columnwidth, clip = true, trim = 0mm 2mm 0mm 0mm]{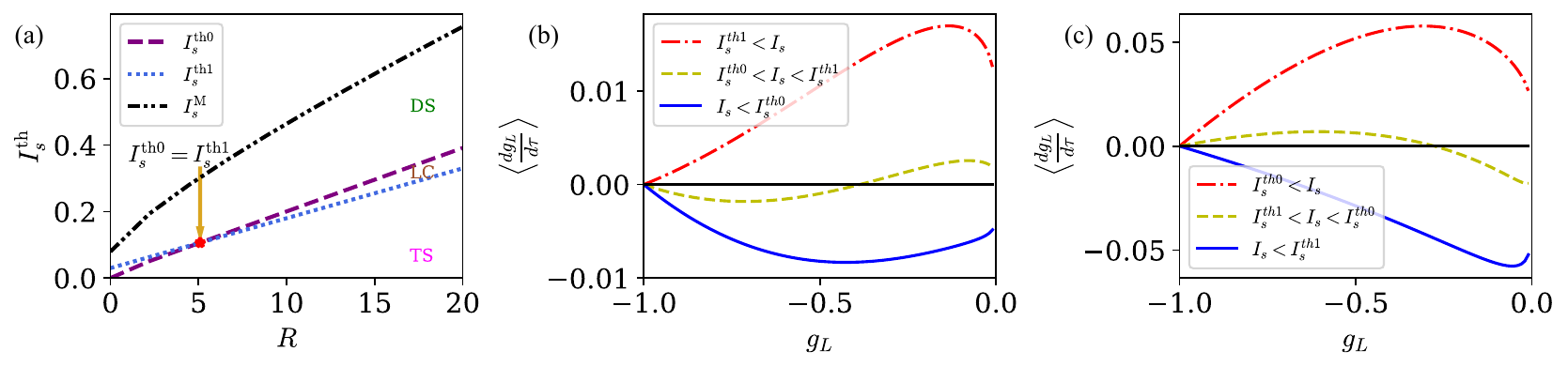}
  \caption{(a) Threshold spin current versus $R$. 
  The asterisk denotes $R = R_{c} = 5.094$ where $I_{s}^\mathrm{th0} = I_{s}^\mathrm{th1}$.  
  For $I_{s}^{\mathrm{thm}} < I_s$, deterministic switching (DS) occurs. Thermal switching region is denoted as TS, \textcolor{black}{while the regime of limit cycles is denoted as LC}. 
  (b) For $R = 3.0$ and \textcolor{black}{$I_{s}^\mathrm{th1} < I_s$}, the rate of change of macrospin energy is positive for $g_L \in (-1,0)$.  
  For $I_{s}^\mathrm{th0} < I_{s} < I_{s}^\mathrm{th1}$, thermal assistance is necessary to cause magnetization reversal. For $I_{s} < I_{s}^\mathrm{th0}$, switching occurs due to thermal noise. (c) For $R = 15.0$ and $I_{s}^\mathrm{th0} < I_{s}$, reversal is deterministic as the rate of energy change is always positive. For $I_{s} < I_{s}^\mathrm{th1}$, thermal switching dominates, while in the intermediate region, thermal assistance is required to cause switching. \textcolor{black}{These results correspond to $\alpha = 0.03$}.}
  \label{fig:ceoa_stuff}
\end{figure*}

Assuming deterministic dynamics and using Eq.~(\ref{avg}) to evaluate for zero energy flow at $g_{L} = 0$ and $g_{L} = -1$ leads to two different threshold currents:~\citep{sun2000spin, pinna2013thermally, taniguchi2013spin} the minimum current required to push the magnetization over the energy barrier into the adjoining basin, $I_{s}^\mathrm{th0} = \alpha \left[\frac{2}{\pi}\sqrt{R\left(1+R\right)}\right]$ and the minimum current required to move the magnetization away from stable equilibrium, $I_{s}^\mathrm{th1} = \alpha[R/2 + 1]$. $R_{c}$ denotes the critical value of $R$ for which $I_{s}^\mathrm{th0}$ equals $I_{s}^\mathrm{th1}$. The threshold currents demarcate regions of deterministic switching from those that require thermal assistance as shown in Fig.~\ref{fig:ceoa_stuff}(a). Figures~\ref{fig:ceoa_stuff}(b) and~\ref{fig:ceoa_stuff}(c) show the rate of change of energy for deterministic dynamics. It is positive for the complete range of macrospin energy only for \textcolor{black}{$I_s^\mathrm{thm} = \mathrm{max}\left(I_s^\mathrm{th0}, I_s^\mathrm{th1}\right) < I_s$} which is consistent with Fig.~\ref{fig:ceoa_stuff}(a). For other values of current, thermal assistance is required for switching.    
The CEOA is valid when the variation in macrospin energy over one precessional cycle is small, i.e. $|T \langle \frac{\partial g_{L}}{\partial \tau} \rangle| \ll \max[|g_{L}|] = 1$. To satisfy this constraint on the variation of macrospin energy, the maximum spin current is given as $I_s^\mathrm{thM} = I_s^\mathrm{th0}\left[1+1/\left(8\alpha \sqrt{R}\right)\right]$.  

\vspace{-10pt}
\section{Analytic switching time model and switching time distributions}\label{sec:analytic}
\vspace{-10pt}
The switching time due to spin torque is defined as the time required for the macrospin energy to change from an initial value $g_{L_{i}}$ to its final value $g_{L_{f}}$. 
\textcolor{black}{Analytic solutions of the switching time presented in this section rely on one or more of the following rational approximations~\citep{rationalApprox2012} for elliptic integrals of Eq.~(\ref{avg})}---
\begin{subequations}
    \begin{gather}
        \begin{split}\label{quad_approx}
            \frac{2}{\pi}\left[E\left(R, g_{L}\right) + g_{L} K\left(R, g_{L}\right)\right] &= A(R) g_{L}^{2} + B(R) g _{L} + C(R),
        \end{split}\\
        \begin{split}\label{K}
            K(x) &= \frac{\pi}{2}\left[\frac{x - 4}{2x - 4}\right],
        \end{split} \\
        \textcolor{black}{\begin{split}\label{E}
            E(x) &= \frac{\pi}{2}\left[1 - \frac{x}{4}\frac{x^{2} - 28x + 64}{4x^{2} - 40x + 64} \right],
        \end{split}}
    \end{gather}
\end{subequations}
where $x = R\frac{\left(1 + g_{L}\right)}{\left(R - g_{L}\right)}$. 
Defining $\tilde{I}_{s} = I_{s}/\alpha$ and \textcolor{black}{using Eqs.~(\ref{quad_approx}) and~(\ref{K}) together in Eq.~(\ref{avg})}, we obtain the magnetization switching time (see Appendix ~\ref{tau_s} for derivation)
\begin{align}\label{tau1}
    \begin{split}
        \tau_{s} &= \frac{1}{4\alpha}\int_{g_{L_{i}}}^{g_{L_{f}}}\left[\frac{3R - g_{L}(R+4)}{R - g_{L}(R+2)}\right]\sqrt{\frac{1+R}{R-g_{L}}} \times \\
    &   \frac{\partial g_{L}}{\left[\tilde{I}_{s}\left(1+g_{L}\right) - \sqrt{1+R}\sqrt{R-g_{L}}\left(Ag_{L}^{2} + B g_{L} + C\right)\right]}.
    \end{split}
\end{align}
In the above equation, the parameters $A$, $B$, and $C$ are functions of $R$ given as $k_{1} + k_{2}R^{k_{3}}$, where the values of $k_{1}$, $k_{2}$ and $k_{3}$ are chosen for different intervals of $R$ to reduce the error in approximating the elliptic integrals. See Table~\ref{tab:coeffs} for details. Note that the values of $A$, $B$, and $C$ are independent of the device geometry and depend only on $R = M_s/H_k$. 
Equation~(\ref{tau1}) is simplified and integrated using partial fractions to arrive at the closed-form expression of switching time:
\begin{align}\label{tauf}
    \begin{split}
    \tau_{s} &= \frac{1}{2\alpha\left(R+2\right)}\vast[\sum_{i = 1}^{5} \frac{N}{D} \log{\left[\frac{(R-g_{L_{f}}) - \lambda_{i}}{(R-g_{L_{i}}) - \lambda_{i}}\right]}\vast. \\
    &+ \left. \sqrt{\frac{R(1+R)}{R+2}} \left\{\frac{\log{\left[\frac{\sqrt{(R-g_{L_{f}})(R+2)} -\sqrt{R(1+R)}}{\sqrt{(R-g_{L_{i}})(R+2)} -\sqrt{R(1+R)}}\right]}}{A\prod\limits_{n=1}^{5} \left(\frac{R(1+R)}{R+2} - \lambda_{i}\right)} \right. \right. \\ 
    &+ \Vast. \left. \frac{\log{\left[\frac{\sqrt{(R-g_{L_{f}})(R+2)} +\sqrt{R(1+R)}}{\sqrt{(R-g_{L_{i}})(R+2)} +\sqrt{R(1+R)}}\right]}}{A\prod\limits_{n=1}^{5} \left(\frac{R(1+R)}{R+2} + \lambda_{i}\right)} \right\} \Vast],
    \end{split}
\end{align}
where $\lambda_{i}$'s are the roots of a fifth-degree
polynomial- $x^{5} - (\frac{B}{A} + 2R)x^{3} + \frac{\tilde{I}_{s}}{A\sqrt{1+R}}x^{2} + (R^{2}+\frac{B}{A}R + \frac{C}{A})x - \frac{\tilde{I}_{s}\sqrt{1+R}}{A}$ with $N = (R+4) + \frac{2R(1+R)}{(R+2)\lambda_{i}^{2} - R(1+R)}$, and $D = 5A\lambda_{i}^{4}-3(B+2AR)\lambda_{i}^{2}+2\frac{\tilde{I}_{s}}{\sqrt{1+R}}\lambda_{i} + (AR^{2} + BR +C)$. 

\begin{table}
\caption{\label{tab:coeffs} Parameters $A$, $B$, and $C$ in Eq.~(\ref{tau1}) are approximated as $k_1 + k_2 R^{k_3}$, where $k_1$, $k_2$, and $k_3$ depend on $R$.}
\begin{ruledtabular}
\begin{tabular}{ccccc}
$R$  &  & $k_{1}$ & $k_{2}$ & $k_{3}$ \\
\hline
\multirow{3}{4em}{[1, 3]} & $A$ & 0.35661 & -0.51244 & -0.38689  \\ 
& $B$ & 1.05148 & -0.55504 & -0.28598 \\ 
& $C$ & 0.61670 &  0.03018 & -1.00153 \\ 
\\
\multirow{3}{4em}{[3, 50]} & $A$ & 0.20223 & -0.38439 & -0.68424  \\ 
&$B$ & 0.81746 & -0.34729 & -0.63939 \\ 
& $C$ & 0.61765 &  0.02994 & -1.08243 \\ 
\\
\multirow{3}{4em}{[50, 100]} & $A$ & 0.17370 & -0.51992 & -0.97986  \\ 
& $B$ & 0.78501 & -0.48295 & -0.97726 \\ 
& $C$ & 0.61755 &  0.02625 & -1.01442 \\ 
\end{tabular}
\end{ruledtabular}
\end{table}
A major advantage of this analytic result is that the approximations of elliptic integrals are independent of the input spin current and Gilbert damping. Additionally, the results obtained in this work are valid for a broad range of $R$ as opposed to prior works~\citep{sun2000spin, vincent2014analytical} that are valid only for $R<R_c$ (= 5.09). 

\textcolor{black}{For the case of $R \rightarrow \infty$, the elliptic integrals in Eq.~(\ref{avg}) are approximated using Eqs.~(\ref{K}) and~(\ref{E})  
This simplifies the energy flow equation, and the switching time is given as}
(see Appendix~\ref{tau_s2} for details)
\begin{align}\label{tau2}
    \begin{split}
        \tau_{s} &= \frac{1}{2\alpha\left[\tilde{I}_{s} - 0.5R\right]}\left[\log{\left[\frac{1+g_{L_{f}}}{1+g_{L_{i}}}\right]}\right. \\
        &+ \left. \frac{b(a-4)(a-8)}{32(a-b)}\log{\left[\frac{1+g_{L_{f}}-a}{1+g_{L_{i}}-a}\right]} \right.  \\
        & -\left.  \frac{a(b-4)(b-8)}{32(a-b)}\log{\left[\frac{1+g_{L_{f}}-b}{1+g_{L_{i}}-b}\right]}\right],
    \end{split}
\end{align}
where $a$ and $b$ are the roots of the quadratic equation $x^{2} - \left(\frac{160\tilde{I}_{s} - 60R}{16\tilde{I}_{s} - 7R}\right)x + \left(\frac{256\tilde{I}_{s} -  128R}{16\tilde{I}_{s} - 7R}\right) = 0$. 

\textcolor{black}{
In the limit of $R\to 0$, the free energy density in Eq. (\ref{energy}) simplifies to $g_{L} = - \left(\mathbf{m} \cdot \mathbf{\hat{n}}_{e}\right)^{2}$, which represents a uniaxial anisotropy ferromagnet. In this case, the switching time is given as}
~\citep{pinna2013spin, pinna2013thermally, sun2000spin, d2019analysis, liu2014dynamics} (see Appendix~\ref{uni} for details)
\begin{align}\label{tau_u}
    \begin{split}
        \tau_{s} &= \frac{1}{2\alpha\left(\tilde{I}_{s}^{2} - 1\right)}\left[\tilde{I}_{s}\left\{\log{\left[\frac{1+\sqrt{-g_{L_{i}}}}{1+\sqrt{-g_{L_{f}}}}\right]} \right. \right.\\
        &\left. \left. - \log{\left[\frac{1-\sqrt{-g_{L_{i}}}}{1-\sqrt{-g_{L_{f}}}}\right]}\right\} - \log{\left[\frac{1 + g_{L_{i}}}{1 + g_{L_{f}}}\right]} \right. \\
        &+ \left. 2\log{\left[\frac{\tilde{I}_{s} - \sqrt{-g_{L_{i}}}}{\tilde{I}_{s} - \sqrt{-g_{L_{f}}}}\right]}\right].
    \end{split}
\end{align}

\begin{figure}[t!]
\centering    
\includegraphics[width = \columnwidth, clip = true, trim = 0mm 0mm 0mm 0mm]{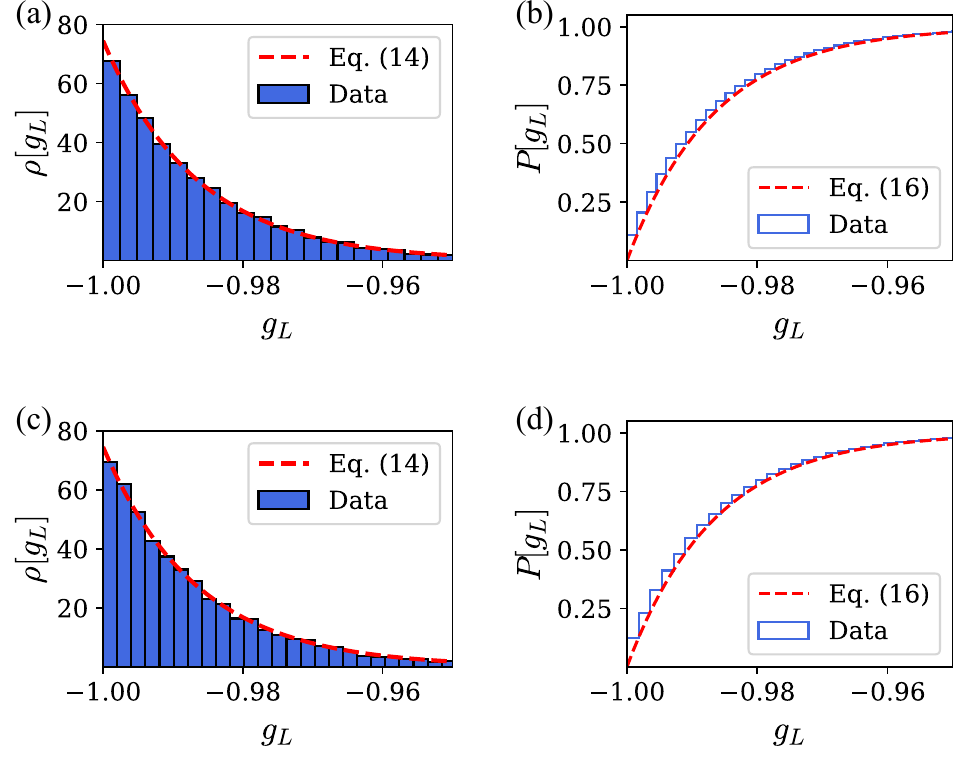}

\caption{\textcolor{black}{Histogram distribution of $g_{L}$ using $\alpha = 0.03$ and $\Delta_{0} = 75$. The top panel results correspond to $R$ = 3.0, while the bottom panel results are for $R$ = 30.0. (a) and (c) denote the PDF, while (b) and (d) denote the CDF.}}\label{equilibrium}
\end{figure}

\begin{figure*}[ht!]
  \centering
  {
  \includegraphics[width = 2\columnwidth, clip = true, trim = 0mm 0mm 0mm 0mm]{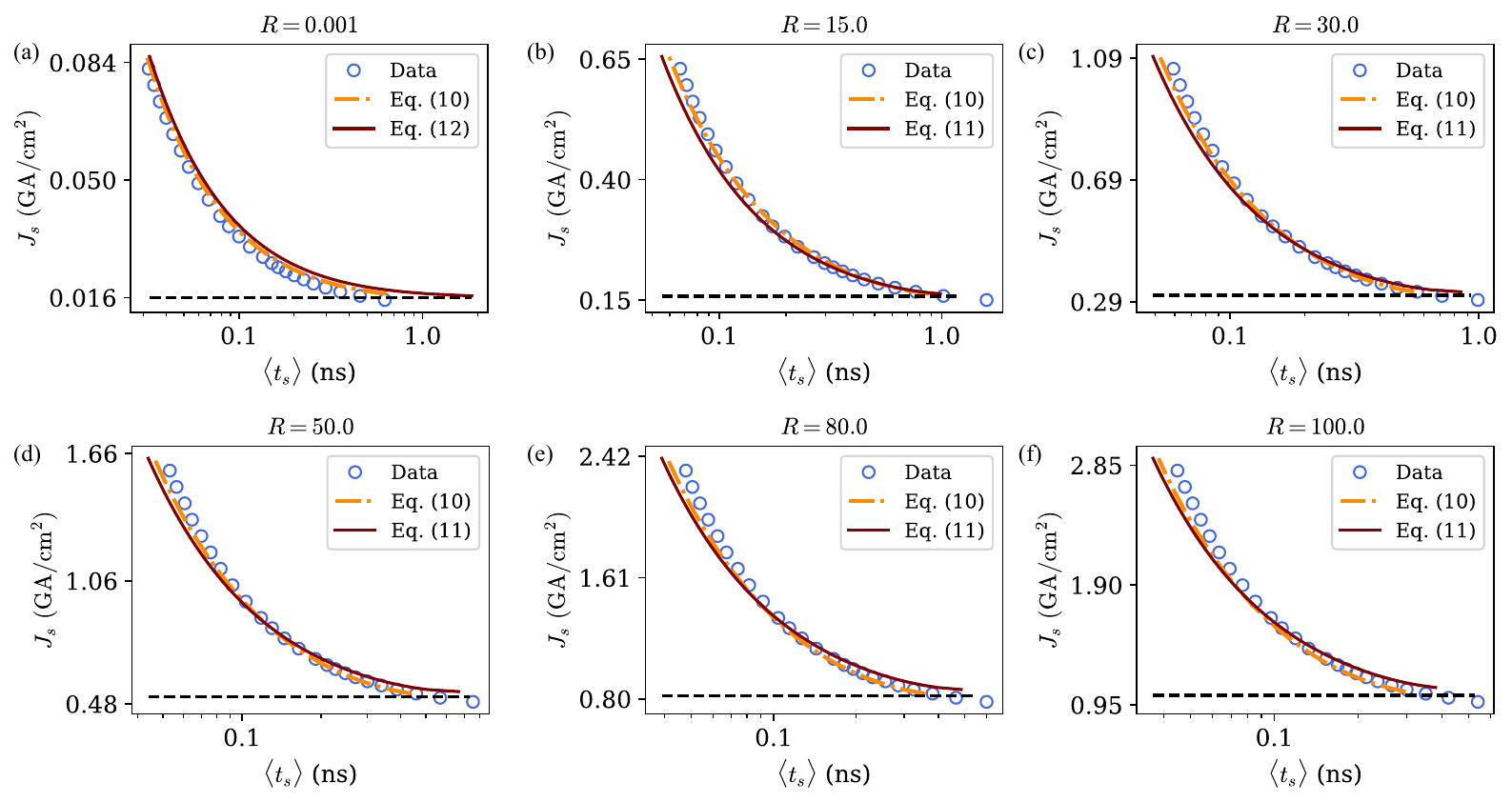}
  }
  \caption{Average switching time $\langle t_{s}\rangle = \frac{\left(1 + \alpha^{2}\right)}{\gamma \mu_{0}} \frac{\langle \tau_{s}\rangle}{H_{k}}$ as a function of injected spin current density, $J_{s}$, for different values of $R$. \textcolor{black}{The horizontal dashed black line in each figure is the threshold current demarcating region of deterministic switching from that of thermal activation.} The numerical data is obtained for an ensemble of $10^{4}$ macrospins.}
  \label{fig:tau_s}
 \end{figure*} 
 
\subsection{\textcolor{black}{Equilibrium Distribution and Average Switching Time}}
In the absence of input spin current, the magnetization is considered to be in thermal equilibrium in its stable energy well. An average switching time, $\langle \tau_{s} \rangle$, is obtained by averaging $\tau_s$ over the equilibrium energy distribution, which in the case of a large energy barrier is the Boltzmann distribution given as
\vspace{-5pt}
\begin{align}\label{equib}
    \begin{split}
        w_{eq}\left(\mathbf{m}\right) &= \frac{1}{Z(\Delta_{0}, R)} \exp\left(-\Delta_{0} g_{L}(\mathbf{m}, 0)\right),   \\
    \end{split}
\end{align}
where $Z(\Delta_{0}, R)$ is the partition function.
\textcolor{black}{We evaluate the Boltzmann distribution function in terms of the free energy random variable} (see Appendix~\ref{rho_gL} for details of the transformation). Accordingly, the probability density function (PDF) is
\begin{equation}\label{equib_energy}
    \textcolor{black}{\rho\left(g_{L}\right) = \frac{1}{Z(\Delta_{0}, R)} \frac{\exp\left(-\Delta_{0} g_{L}\right)}{\sqrt{-g_{L}}}},
\end{equation}
where the partition function
\begin{align}\label{partition}
    \begin{split}
        Z(\Delta_{0}, R) &= \int_{-1}^{0} \rho(x) dx = \frac{2\exp\left(\Delta_{0}\right)F(\sqrt{\Delta_{0}})}{\sqrt{\Delta_{0}}}. 
    \end{split}
\end{align}
\textcolor{black}{The cumulative distribution function (CDF)} is
\begin{align}\label{cdf}
    \begin{split}
        \textcolor{black}{P\left(g_{L}\right)} &= \textcolor{black}{\int_{-1}^{g_{L}} \rho(x) dx} \\
        &= \textcolor{black}{1 - \exp\left(-\Delta_{0}\left(1+g_{L}\right)\right)\frac{F\left(\sqrt{-\Delta_{0} g_{L}}\right)}{F\left(\sqrt{\Delta_0}\right)}},       
    \end{split}
\end{align}
where $F(x) = \exp(-x^{2})\int_{0}^{x} \exp(y^{2}) dy$ is the Dawson's integral.

The magnetization is considered to have switched successfully when it crosses the separatrix $\left(g_{L_{f}} = 0\right)$ and consequently moves into the adjoining energy well. 
Once the magnetization moves into the target energy well, the spin current could be switched off. The magnetization would eventually settle into its stable well due to its intrinsic damping. Therefore, $\langle \tau_{s} \rangle$ is given as 
\begin{equation}
    \langle\tau_{s}\rangle = \int_{-1}^{0} \rho\left(g_{L_{i}}\right) \ \tau_{s}\left(g_{L_{i}}, g_{L_{f}} = 0\right) \ d g_{L_{i}}.
\end{equation}

\subsection{\textcolor{black}{Distribution Functions and Write Error Rate}}
\vspace{-10pt}
Defining a random variable switching time $T_{s}$, the fraction of macrospins in an ensemble that have switched from an anti-parallel to a parallel state at time $\tau_s$ is

\begin{flalign}\label{cdf_tau}
    \begin{split}
        &P_{T_{s}}\left[\tau_{s}\right] = 1 - \int_{-1}^{g_{L_{i}}(\tau_{s})}\frac{1}{Z(\Delta_{0}, R)} \frac{\exp\left(-\Delta_{0} x\right)}{\sqrt{-x}} dx \\
        & = \exp\left(-\Delta_{0}\left(1+g_{L_{i}}(\tau_{s})\right)\right)\frac{F\left(\sqrt{-\Delta_{0}g_{L_{i}}(\tau_{s}})\right)}{F\left(\sqrt{\Delta_0}\right)},  
    \end{split}
\end{flalign}
where $g_{L_{i}}(\tau_s)$ corresponds to the initial energy for switching time $\tau_{s}$. An exact analytic expression for $g_{L_{i}}(\tau_{s})$ is not feasible; however, the monotonically decreasing nature of Eqs.~(\ref{tauf})-(\ref{tau_u}) makes it possible to numerically invert~\citep{inverse} them. 
$P_{T_{s}}\left[\tau_{s}\right]$ includes all macrospins in an ensemble with $g_{L} \geq g_{L_{i}}(\tau_{s})$ as we consider only deterministic switching in this work.
$P_{T_{s}}\left[\tau_{s}\right]$ also refers to the probability of switching of a macrospin with $\tau_{s} \geq T_{s} $; therefore, $P_{T_{s}}\left[\tau_{s}\right] = Pr\left[T_{s} \leq \tau_{s}\right]$.
Finally, the probability density function of the switching time is    
\begin{align}\label{pdf_tau}
    \begin{split}
        \rho_{T_{s}}(\tau_{s}) &= \frac{d P_{T_{s}}\left[\tau_{s}\right]}{d \tau_{s}} \\         
        & = \frac{\sqrt{\Delta_{0}}}{2 F(\sqrt{\Delta_0})} \frac{\exp\left(-\Delta_{0}(1 + g_{L_{i}}(\tau_{s}))\right)}{\sqrt{-g_{L_{i}}(\tau_{s})}} \ \left|\frac{d g_{L_{i}}(\tau_{s})}{d \tau_{s}}\right|. \\
    \end{split}
\end{align}

The write-error rate (WER) quantifies the probability of unsuccessful spin torque switching of the magnet. Using Eq.~(\ref{cdf_tau}), the WER is 
\begin{align}\label{wer}
    \begin{split}
        \text{WER} &= 1 - P_{T_{s}}[\tau]\\
        & = 1 - \exp\left(-\Delta_{0}\left(1+g_{L_{i}}(\tau_{s})\right)\right)\frac{F\left(\sqrt{-\Delta_{0}g_{L_{i}}(\tau_{s})}\right)}{F\left(\sqrt{\Delta_{0}}\right)}. 
    \end{split}
\end{align}

\begin{figure}[ht!]
  \centering
  {
  \includegraphics[width = \columnwidth, clip = true, trim = 0mm 0mm 0mm 0mm]{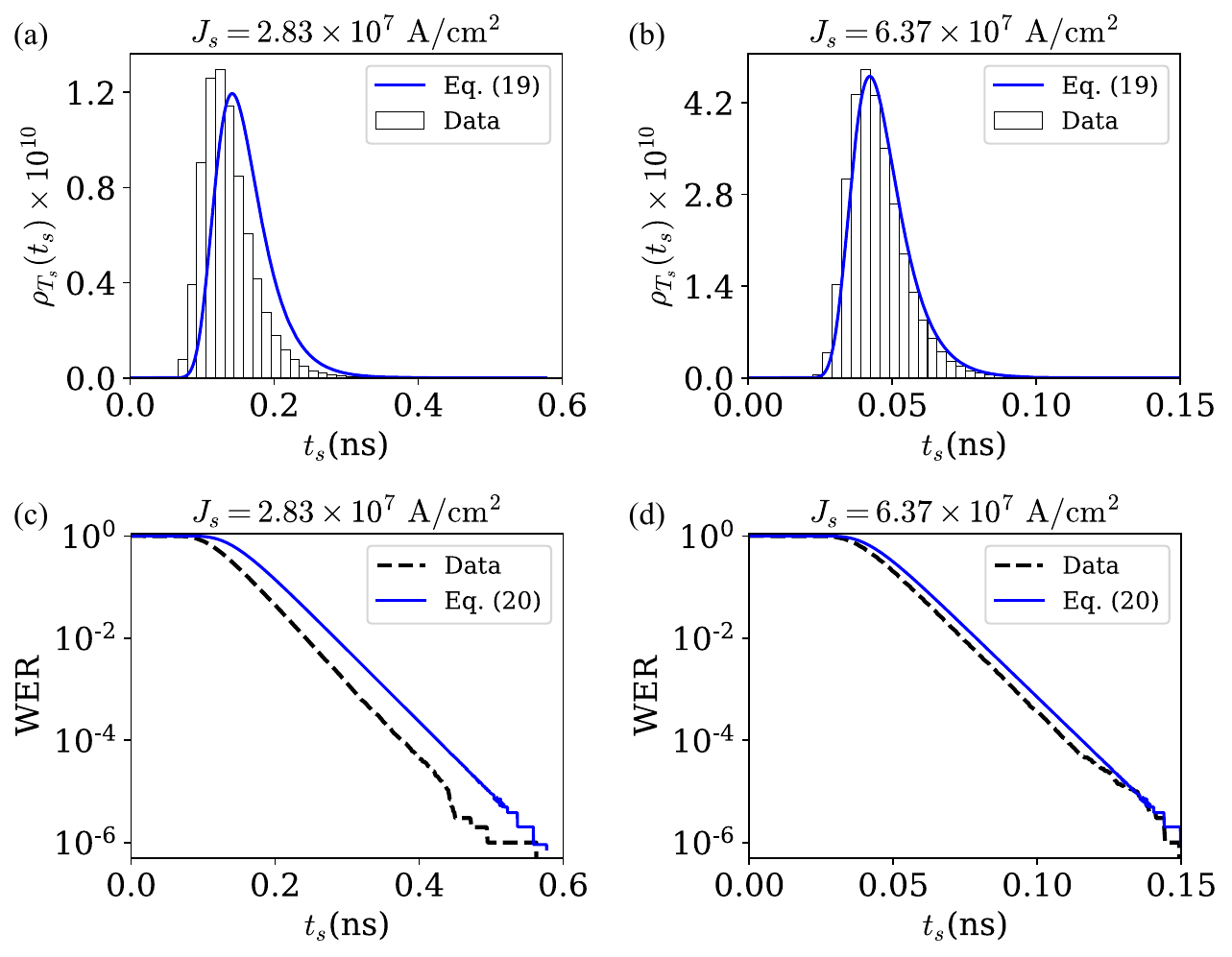}
  }
  \caption{All results reported for $R$ = 0.001 (uniaxial anisotropy) and $J_s^\mathrm{th1} = 1.6\times 10^7$ A/cm$^2$. \textcolor{black}{The top panel shows the PDF, while the bottom panel shows the WER.} The accuracy of analytical results improves as the input spin current density increases with respect to $J_s^\mathrm{th1}$.}
  \label{fig:R0_pdf_wer}
 \end{figure}

\begin{figure}[ht!]
  \centering
  {
  \includegraphics[width = \columnwidth, clip = true, trim = 0mm 0mm 0mm 0mm]{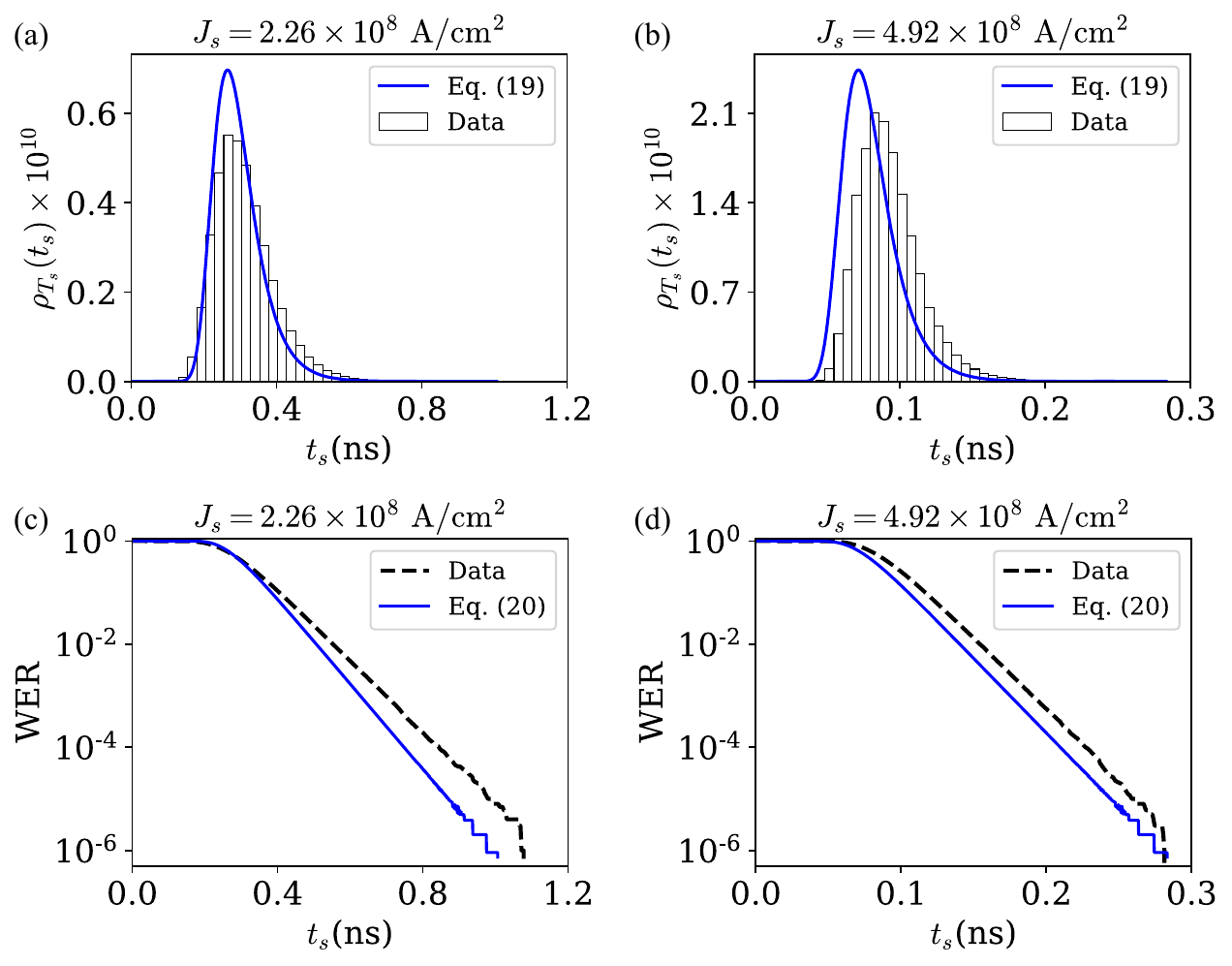}
  }
  \caption{All results are reported for $R$ = 15 and $J_s^\mathrm{th0} = 1.58\times 10^8$ A/cm$^2$. \textcolor{black}{The top panel shows the PDF, while the bottom panel shows the WER.} The accuracy of analytical solutions improves for spin current densities larger than the threshold current density. \textcolor{black}{Though CEOA is strictly valid for $J_s < J_s^\mathrm{thM} = 2.1J_{s}^\mathrm{th0}$, the analytic model agrees well with the numerical solution for current larger than $J_s^\mathrm{thM}$.} }
  \label{fig:R15_pdf_wer}
\end{figure}

\vspace{-5pt}
\subsection{\textcolor{black}{Model Validation}}
\vspace{-5pt}
To benchmark our analytic results, we solve Eq.~(\ref{sLLGS}) numerically using the Heun integration scheme implemented in CUDA and run in parallel on GPUs. Numerical simulations were calibrated against published results to ensure their accuracy.~\citep{pinna2013thermally, pinna2013spin, kani2016probability} For all numerical simulations presented in this work, the time step of integration was set as 0.3 ps. \textcolor{black}{The Gilbert damping constant of the free-layer is taken as $\alpha=0.03$, while its cross-sectional area is assumed to be $Ar = \pi/4 \times 30 \times 15 \ \mathrm{nm}^{2}$.
Different values of $R$, each signifying a different ferromagnetic material with its respective $M_{s}$ and $K_{u}$ values, are considered. 
The thickness of the free layer is adjusted to achieve an energy barrier $\Delta_{0} = 75$.~\citep{chun2012scaling} 
Simulation results corresponding to values of $\alpha$ other than 0.03 are not reported for brevity, as the key features and trends of switching dynamics remain the same. 
Finally, a wide range of $R$ values is selected to show the applicability and robustness of the model.}

\begin{figure}[ht!]
  \centering
  {
  \includegraphics[width = \columnwidth, clip = true, trim = 0mm 0mm 0mm 0mm]{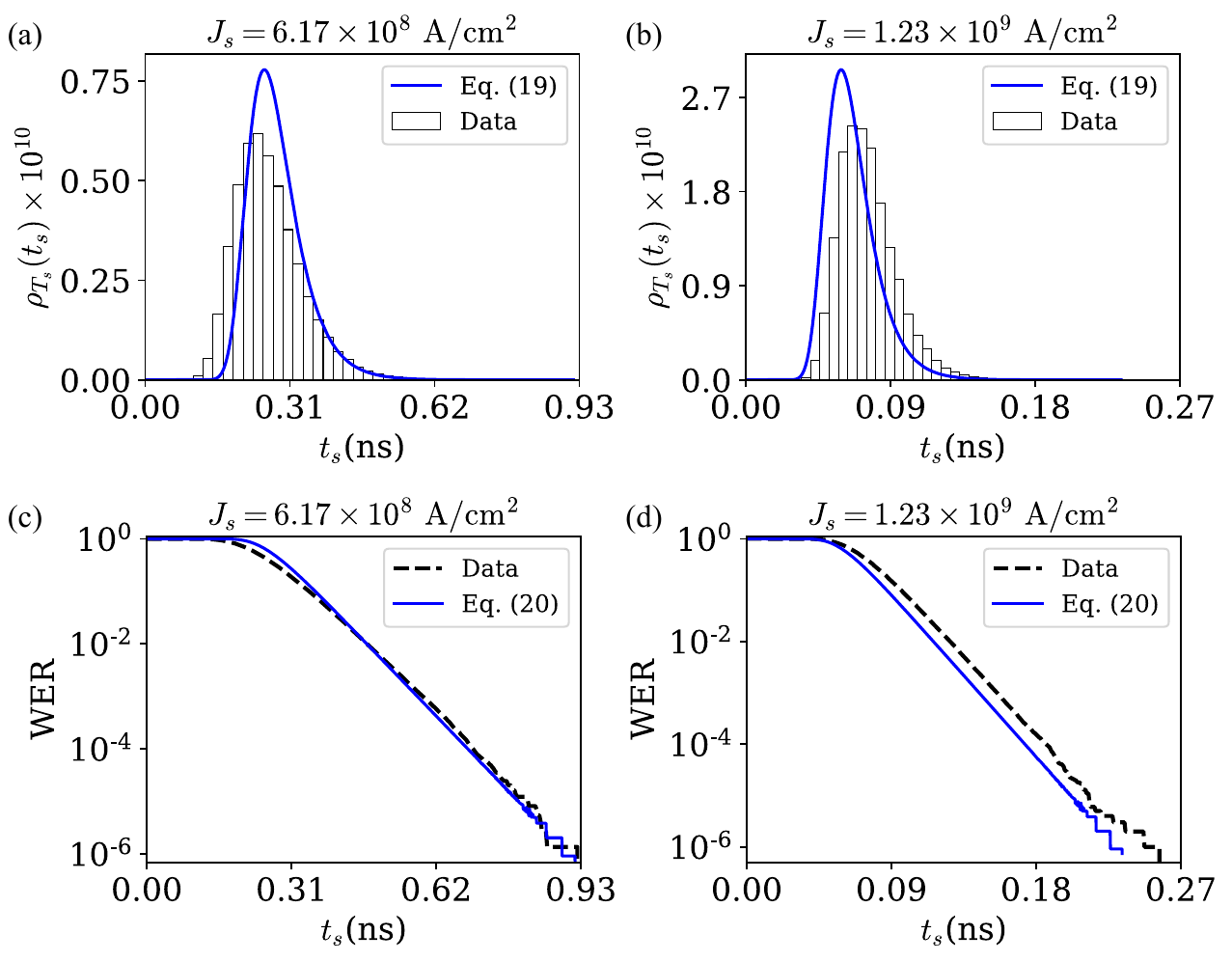}
  }
  \caption{All results are reported for $R$ = 50 and $J_s^\mathrm{th0} = 5.14\times 10^8$ A/cm$^2$. \textcolor{black}{The top panel shows the PDF, while the bottom panel shows the WER. The accuracy of analytic results improves for spin current density larger than the threshold current density. The analytic model is fairly accurate even for $J_s > J_s^\mathrm{thM} = 1.6J_s^\mathrm{th0}$ showing the robustness of our model.}}
  \label{fig:R50_pdf_wer}
\end{figure} 

Figure~\ref{equilibrium} shows that Eqs.~(\ref{equib_energy}) and~(\ref{cdf}) describe the equilibrium numerical distribution very well.

\vspace{-5pt}
\section{{Results and Discussion}}\label{sec:results_disc}
\vspace{-10pt}
\textcolor{black}{Figure~\ref{fig:tau_s} shows the average switching time of an ensemble of $10^{4}$ independent macrospins},  $\langle t_{s}\rangle = \frac{\left(1 + \alpha^{2}\right)}{\gamma \mu_{0}} \frac{\langle \tau_{s}\rangle}{H_{k}}$ as a function of the input spin current density, \textcolor{black}{$J_{s} = \Delta_{0} (4e k_{B}T/\hbar) (I_{s}/Ar)$}, where $e$ is the electron charge, and $\hbar$ is the reduced Planck's constant. 
\textcolor{black}{For $R = 0.001$ (uniaxial anisotropy), there is an excellent agreement between numerical data and closed-form solutions given in Eqs.~(\ref{tauf}) and~(\ref{tau_u}) for moderate to large current levels.} However, for current levels approaching the threshold switching currents, numerical results predict a lower average switching time as the presence of thermal noise aids the switching process.   

\textcolor{black}{For $R \gg 1$ (biaxial anisotropy)}, $\langle t_{s}\rangle$ obtained using Eq.~(\ref{tau2}) predicts a larger average switching time compared to the numerical results for current levels comparable to the threshold value. This is expected since the analytic solutions neglect the effect of thermal noise during the switching process. 
As the input spin current density ($J_s$) increases beyond the threshold value, \textcolor{black}{$J_s^\mathrm{th0}$ ($ = \Delta_{0} (4e k_{B}T/\hbar) (I_s^\mathrm{th0}/Ar)$)}, the agreement between analytic and numerical results improves. 
Near the threshold current level, the average switching time obtained from Eq.~(\ref{tauf}) is slightly lower than that obtained from numerical results. This slight deviation is due to the quadratic approximation of elliptic integrals in Eq.~(\ref{avg}).  
\textcolor{black}{Note that for $J_s > J_s^\mathrm{thM} ( = \Delta_{0} (4e k_{B}T/\hbar) (I_s^\mathrm{thM}/Ar))$, the validity of CEOA is not fully justified as the average rate of energy change becomes large.~\citep{pinna2013thermally}} 
\textcolor{black}{However, the general trend of switching time stays the same as the error due to CEOA does not increase abruptly but rather increases slowly. We observe that the error between the switching time from solution of Eqs.~(\ref{sLLGS}) and~(\ref{tauf}) is between $5 - 12\%$ for different values of $R$ for current as large as $2J_s^\mathrm{thM}$. }

\begin{figure}[ht!]
  \centering
  {
  \includegraphics[width = \columnwidth, clip = true, trim = 0mm 0mm 0mm 0mm]{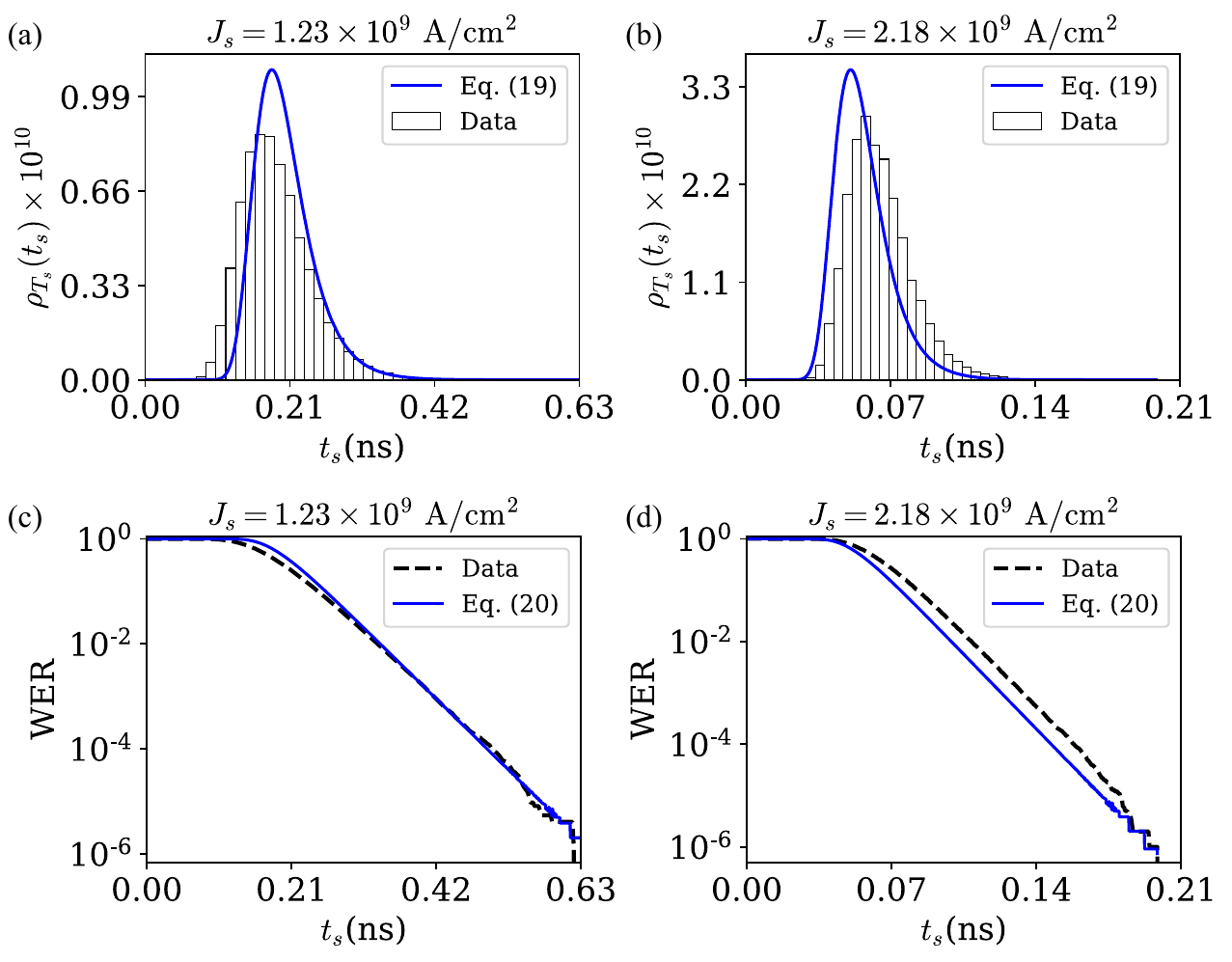}
  }
  \caption{All results are reported for $R$ = 100 and $J_s^\mathrm{th0} = 1.02\times 10^9$ A/cm$^2$. \textcolor{black}{The top panel shows the PDF, while the bottom panel shows the WER. The accuracy of analytical results improves as $J_s$ increases above $J_s^\mathrm{th0}$. For $J_s > J_{s}^\mathrm{thM} = 1.42J_{s}^\mathrm{th0}$, the accuracy of the analytic model reduces but continues to predict the numerical results with good accuracy. }}
  \label{fig:R100_pdf_wer}
\end{figure}

\textcolor{black}{The PDF and WER obtained using analytic results of Eqs.~(\ref{pdf_tau}) and (\ref{wer}) are compared against numerical solution in Figs.~\ref{fig:R0_pdf_wer}-\ref{fig:R100_pdf_wer} for an ensemble of $10^6$ macrospins.
It is observed that for the uniaxial case, the accuracy of the PDF and WER improves as the applied spin current increases.~\citep{kani2016probability}} For $R \geq 15$, the accuracy of analytic solutions also increases as current increases from $J_s^\mathrm{th0}$ toward $J_{s}^\mathrm{thM}$. \textcolor{black}{Though for spin currents larger than $J_{s}^\mathrm{thM}$, the accuracy of analytic results reduces as the validity of CEOA becomes questionable, the model is fairly robust and predicts numerical results well.} To arrive at analytic results reported in Figs.~\ref{fig:R15_pdf_wer}-\ref{fig:R100_pdf_wer}, we numerically invert Eq.~(\ref{tau2}) due to its simplicity.
\begin{table}[t!]
\caption{\label{tab:materials} \textcolor{black}{List of common ferromagnetic materials, their saturation magnetization $M_{s}$, uniaxial anisotropy energy density $K_{u}$ and damping constant $\alpha$.}}
\begin{ruledtabular}
\begin{tabular}{lccccr}
Materials  &  $M_{s}$ (T)  & $K_{u}$ ($\mathrm{MJ/m^{3}}$) & $R$ & $\alpha$ & Ref.\\
\hline
Terfenol-D & 1.0 & 0.39 & 1.04 & 0.1 & \citen{kani2017modeling} \\
Co  & 1.81 & 0.41 & 3.2 & 0.02 & \citen{coey2010magnetism}\\
$\mathrm{Co_{0.6}Fe_{0.2}B_{0.2}}$ & $1.2$ & 0.095 & 6.0 & 0.015 & \citen{rowlands2011deep, yakata2009influence} \\
NiMnSb & 0.84 & 0.013 & 21.6 &0.002 &\citen{coey2010magnetism, durrenfeld2015spin}\\
Fe  & 2.15 & 0.048& 38.3 & 0.001 & \citen{coey2010magnetism, yoda2010high}\\
EuO & 2.36 & 0.044 & 50.4 & 0.015 & \citen{coey2010magnetism}\\
FeGaB & 1.63 & 0.0198 & 53.4 & 0.1 & \citen{kani2017modeling}\\
\end{tabular}
\end{ruledtabular}
\end{table}

\begin{figure}[ht!]
  \centering
  {
  \includegraphics[width = \columnwidth, clip = true, trim = 0mm 0mm 0mm 0mm]{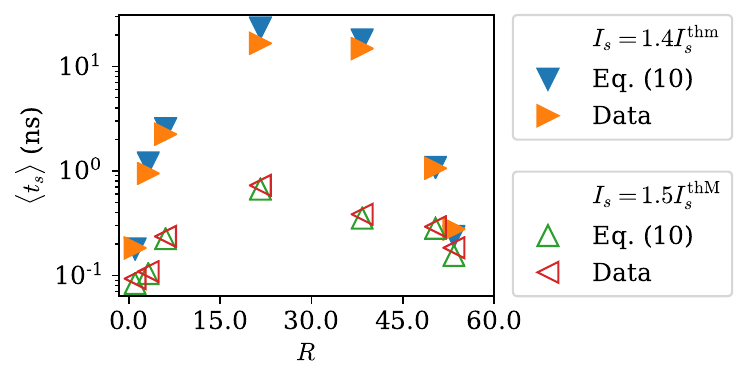}
  }
  \caption{\textcolor{black}{Average switching time for materials listed in Table~(\ref{tab:materials}) for two different currents, viz. $I_{s} = 1.4I_{s}^{\mathrm{thm}}$ and $I_{s} = 1.5 I_{s}^{\mathrm{thM}}$.}}
  \label{fig:R}
\end{figure}

\textcolor{black}{Figure~\ref{fig:R} shows the average switching time for a few 
ferromagnetic thin films.
Both numerical and analytic average switching times (Eqs.~(\ref{sLLGS}) and~(\ref{tauf})) are for the material parameters listed in Table~\ref{tab:materials}. The larger value of spin current is chosen to demonstrate applicability of the model
beyond the strict validity of CEOA. As expected, the average switching time decreases for larger value of spin current. Also, for two materials with similar values of $R$, the switching time is lower for the material with a higher damping constant.}

\vspace{-5pt}
\subsection{Applications of spin torque memory in the deterministic and thermally activated regime}
\vspace{-5pt}
\textcolor{black}{Conventional artificial intelligence (AI) hardware accelerators based on von Neumann architectures suffer from significant power dissipation and communication overheads.~\cite{nagasaka2010statistical, abadi2016tensorflow, indiveri2015memory} The introduction of non-volatile magnetic memory in a hybrid magnetic-silicon hardware could lead to lower power dissipation even for computationally expensive machine learning tasks that routinely process large-scale datasets.~\citep{upadhyay2019emerging}  
Neuromorphic architectures that co-locate memory and compute elements are well suited to 
leverage the non-volatile spin-torque memory discussed in this work. Such memory can be used as a cache memory in general-purpose processors and for storing the weights pertinent to machine learning tasks in a neuromorphic architecture.~\citep{grollier2016spintronic}} 

\textcolor{black}{
Beyond serving as non-volatile memory elements, the device architectures illustrated in Fig.~\ref{memory} can also function as stochastic oscillators when operated in the thermally activated regime for which $I_{s} < I_{s}^{\mathrm{thm}}$. In this regime, the input spin current controls the rate at which the magnetization of the ferromagnetic thin film fluctuates between its two stable states.~\cite{camsari2017implementing} Several concepts of neuromorphic computing have been proposed in the literature based on fluctuating magnets.~\citep{vincent2015spin, torrejon2017neuromorphic, grollier2016spintronic, sengupta2017encoding} 
Moreover, by relaxing the constraint on the energy barrier, the thresholding behavior of mono-domain ferromagnets can mimic neural dynamics in an energy efficient manner.
It has also been shown that the structures of Fig.~\ref{memory} can implement synaptic functionality,~\cite{vincent2015spin, yue2019brain} which is necessary for weight storage and update in neuromorphic hardware.    
Energy efficient probabilistic computing for image processing and machine learning,~\citep{rangarajan2017energy} and spin logic~\citep{venkatesan2015spintastic, camsari2017stochastic} are few other possible applications in the thermally activated regime. }


\textcolor{black}{The thermally activated regime is also important from the point of view of studying the fundamental scaling laws of spin torque driven dynamics of ferromagnetic materials. The scaling of the energy barrier height of ferromagnets with spin current is typically modeled as $\Delta = \Delta_{0} \left(1 - I_{s}/I_{s}^{\mathrm{th1}}\right)^{\xi}$.
The value of $\xi$ in the limit of $I_{s} \ll I_{s}^{\mathrm{th1}}$ has been a subject of debate for over two decades.~\citep{apalkov2005spin, li2004thermally, taniguchi2011thermally, pinna2012thermally, taniguchi2013thermally} Recently, it was theoretically postulated~\citep{taniguchi2011thermally, pinna2012thermally} that $\xi = 2$ for the case of uniaxial anisotropy ($R \to 0$). For the biaxial case ($R \gg 1$), however, the value of $\xi$ was estimated to be $2.2$ by Taniguchi and coworkers based on the Fokker-Plank (FP) equation.~\citep{taniguchi2013thermally} Pinna and coworkers note the current dependence of $\xi$ via the the principle of action~\citep{pinna2012thermally, pinna2013thermally} on Eq.~(\ref{avg}).
In our opinion, it might be possible to construct a FP representation from Eq.~(\ref{avg}) and solve it with the rational approximations Eqs.~(\ref{K})-(\ref{E}) to arrive at an analytical expression for $\xi$. However, full analysis of dynamics in the thermally activated regime is out of scope of this work.}

\vspace{-5pt}
\section{Conclusion}\label{sec:conclusion}
\vspace{-10pt}
Analytic models of average switching time, probability distribution function of switching times, and the write-error rate developed in this paper for thin-film magnets with biaxial anisotropy show good agreement against numerical results for moderate to large spin current densities. 
In the vicinity of the threshold spin current density, the error between analytic and numerical data is significant due to thermal noise. 
For very large spin current densities, the constant energy orbit averaging approach adopted in this work becomes inadequate, even though the error between numerical and analytic results is well under a tolerance limit.
The models of this paper should complement experimental results and aid the analysis, design and development of non-volatile memory driven by both spin-transfer and spin-orbit torques.

\section*{acknowledgements}

This work was supported partially by the Semiconductor Research Corporation and the National Science Foundation (NSF) through ECCS 1740136.

\appendix
\vspace{-5pt}
\section{Simplifying Eq.~(\ref{gdot})}\label{derive_gdot}
\vspace{-10pt}
Without any loss of generality, we consider the easy and hard axes to coincide with the $\mathbf{\hat{x}}$ and $\mathbf{\hat{z}}$ axes, respectively. Next, we consider spin polarization of the fixed layer to be in the plane of the magnet at an angle $\phi$ with the easy axis, therefore, $\mathbf{\hat{n}}_{p} = \cos{\phi} \ \mathbf{\hat{x}} + \sin{\phi} \ \mathbf{\hat{y}}$. Therefore, for zero external magnetic field ($\mathbf{H}_{a} = 0$), the effective magnetic field $\mathbf{h}_{\mathrm{eff}} = m_{x} \mathbf{\hat{x}} - R m_{z} \mathbf{\hat{z}} + \mathbf{h}_{T}$, where $\mathbf{h}_{T}$ is the thermal field. Therefore, the three components of Eq.~(\ref{sLLGS}) are
\begin{subequations}
    \begin{flalign}\label{mx_dotS}
    \begin{split}
        \frac{\partial {m}_{x}}{\partial \tau} &= R m_{y} m_{z} + \alpha m_{x} \left( 1 - m_{x}^{2} + R m_{z}^{2} \right) \\
        &+ I_{s}\left( \cos{\phi} - m_{x}\left(\cos{\phi} \ m_{x} + \sin{\phi}\ m_{y}\right)\right)
         \\
        & - \alpha I_{s} \ m_{z}\sin{\phi} \\
        & + n_{S, x},
    \end{split}
\end{flalign}
\begin{flalign}\label{my_dotS}
    \begin{split}
        \frac{\partial {m}_{y}}{\partial \tau} &= -\left(R + 1\right) m_{x} m_{z} - \alpha m_{y} \left( m_{x}^{2} - R m_{z}^{2} \right)  \\
        &+ I_{s} \left(\sin{\phi}- m_{y}\left(\cos{\phi} \ m_{x} + \sin{\phi}\ m_{y}\right)\right)\\
        & + \alpha I_{s} \ m_{z} \cos{\phi}  \\
        &+ n_{S, y},
    \end{split}
\end{flalign}
\text{and}
\begin{flalign}\label{mz_dotS}
    \begin{split}
        \frac{\partial {m}_{z}}{\partial \tau} &= \ m_{x} m_{y} \ - \ \alpha m_{z} \left(R +  m_{x}^{2} \ -\  R m_{z}^{2}\right) \\
        &- I_{s} m_{z}\left(\cos{\phi} \ m_{x} \ + \  \sin{\phi}\ m_{y} \right) \\
        &+ \alpha I_{s} \left( \sin{\phi}\ m_{x} \ - \ \cos{\phi} \ m_{y} \right)  \\
        &+ n_{S, z},
    \end{split}
\end{flalign}
\end{subequations}
where each of $n_{S, p}$ denotes thermal noise component in the Stratonovich sense. \textcolor{black}{From Eqs.~(\ref{thermal}) and~(\ref{diffusivity}) we have} 
\begin{flalign}\label{noise}
    \begin{split}
        &\begin{pmatrix}
           n_{S, x}\\
            n_{S, y}\\
            n_{S, z}
        \end{pmatrix} = \mathcal{D}
                \circ \begin{pmatrix}
                        h_{T, x}\\
                        h_{T, y}\\
                        h_{T, z}
                    \end{pmatrix}  = \sqrt{D} \ \mathcal{D}
                \circ \begin{pmatrix}
                        \dot{W}_{x}\\
                        \dot{W}_{y}\\
                        \dot{W}_{z}
                    \end{pmatrix},
    \end{split}
\end{flalign}
where $\mathbf{\dot{W}}$ represents a 3D stochastic Wiener process whose each component is a Gaussian random variable with zero mean and unit standard deviation. $\mathcal{D}$ is referred to as the diffusion matrix and is given as
\begin{equation*}
    \begin{pmatrix}
                \alpha\left(1 - m_{x}^{2}\right) & m_{z} - \alpha m_{x} m_{y} & -m_{y} - \alpha m_{x} m_{z} \\
                -m_{z} - \alpha m_{x} m_{y} & \alpha\left(1 - m_{y}^{2}\right) & m_{x} -\alpha m_{y} m_{z}\\
                m_{y}-\alpha m_{x} m_{z} & -m_{x} - \alpha m_{y} m_{z} & \alpha\left(1 - m_{z}^{2}\right)
        \end{pmatrix}.
\end{equation*}
Substituting Eqs.~(\ref{mx_dotS}) and~(\ref{mz_dotS}) into Eq.~(\ref{gdot}) leads to
\begin{subequations}
    \begin{align}\label{gdot3}
    \begin{split}
        \frac{\partial g_{L}}{\partial \tau} &= 2 R m_{z} \left[m_{x} m_{y} - \alpha m_{z} \left(R - g_{L}\right) \right.\\
        & - I_{s} m_{z}\left(\cos{\phi} \ m_{x} + \sin{\phi}\ m_{y}\right)\\
        &+ \left. \alpha I_{s} \left(\sin{\phi}\ m_{x} - \cos{\phi} \ m_{y} \right) \right]\\
        & - 2 m_{x} \left[R m_{y} m_{z} + \alpha m_{x} \left( 1 + g_{L} \right) \right. \\  
        & +  I_{s} \left( \cos{\phi} - m_{x} \left(\cos{\phi} \ m_{x} + \sin{\phi} \ m_{y}\right) \right) \\
        & - \left. \alpha I_{s} \sin{\phi}\ m_{y} \right] \\
        &+ 2 \sqrt{D} \begin{pmatrix}
                        d_{x}  & d_{y} & d_{z}
                    \end{pmatrix} \circ \begin{pmatrix}
                        \dot{W}_{x}\\
                        \dot{W}_{y}\\
                        \dot{W}_{z}
                    \end{pmatrix};  \\
    \end{split}\\
    \begin{split}
\mathrm{and} &      \begin{pmatrix}
                        d_{x}  \\ d_{y} \\ d_{z}
                    \end{pmatrix} = 
                        \begin{pmatrix}
                        R m_{z}\mathcal{D}_{31} - m_{x} \mathcal{D}_{11}  \\ R m_{z} \mathcal{D}_{32} - m_{x} \mathcal{D}_{12} \\ R m_{z} \mathcal{D}_{33}- m_{x} \mathcal{D}_{13}
                    \end{pmatrix}.
    \end{split}
\end{align}
\end{subequations}
It can be observed from Eq.~(\ref{gdot3}) that each of damping, spin current and thermal noise contribute to the rate of energy change. In addition, the last term in Eq.~(\ref{gdot3}) is simplified to $\sqrt{d_{x}^{2} + d_{y}^{2} + d_{z}^{2}}\sqrt{\dot{W}_{x}^{2} + \dot{W}_{y}^{2} + \dot{W}_{Z}^{2}}$ which leads to
\begin{align}\label{gdot4}
    \begin{split}
        \frac{\partial g_{L}}{\partial \tau} &= -2 \alpha \left[\left(1+R\right)m_{x}^{2} + g_{L}\left(R - g_{L}\right)\right. \\
        &+ \frac{I_{s}}{\alpha}\left(\left(1 + g_{L}\right) \cos{\phi} \ m_{x} + g_{L} \sin{\phi} \ m_{y}\right) \\
        & \left. + I_{s} \left( \sin{\phi}\ m_{x} m_{y} - R\left(\sin{\phi} \ m_{x} m_{z} - \cos{\phi} \ m_{y} m_{z} \right)\right) \right] \\
        &+ 2\sqrt{\frac{\alpha}{\Delta_{0}}}\sqrt{\left(1+R\right)m_{x}^{2} + g_{L}\left(R - g_{L}\right)} \circ \dot{W}_{g_{L}},
    \end{split}
\end{align}
where $\dot{W}_{g_{L}}$ is 1D white Gaussian noise and it acts away from the constant-energy orbit along a normal.~\citep{pinna2014spin}

Now average $\frac{\partial g_{L}}{\partial \tau}$ over a constant-energy orbit as the non-conservative effects act over long time-scales whereas we are more interested in studying their effects on rapid periodic motion. In the limit of zero damping, zero spin current at absolute zero temperature, the constant energy solutions to Eqs.~(\ref{mx_dotS})-(\ref{mz_dotS}) in the anti-parallel well are given as~\citep{mayergoyz2009nonlinear,pinna2013thermally,pinna2014spin} 
\begin{subequations}
    \begin{flalign}
        \begin{split}
            &m_{x}^{c}(t) = - \sqrt{\frac{R - g_{L}}{R + 1}} \ dn\left[\sqrt{R - g_{L}} \ t, k^{2}\right], 
        \end{split} \\
        \begin{split}
            &m_{y}^{c}(t) = \sqrt{1 + g_{L}} \ sn\left[\sqrt{R-g_{L}}\ t, k^{2}\right],
        \end{split}\\
        \begin{split}
            &m_{z}^{c}(t) = \sqrt{\frac{1 + g_{L}}{R + 1}} \ cn\left[\sqrt{R - g_{L}} \ t, k^{2}\right],
        \end{split}
    \end{flalign}
\end{subequations}
where $k^{2} \equiv R\frac{1 + g_{L}}{R - g_{L}}$, $0 < k^{2} < 1$ and $sn[\cdot]$, $cn[\cdot]$, $dn[\cdot]$ are Jacobi elliptic functions. Magnetization oscillates around the easy axis with a time period, $T\left(g_{L}\right) = \frac{4}{\sqrt{R - g_{L}}} K\left(k^{2}\right)$, where $K\left(k^{2}\right)$ is the complete elliptic integral of first kind.
In Eq.~(\ref{gdot4}) only the time averages of $m_{x}$ and $m_{x}^{2}$ are non-zero while those of $m_{y}$, $m_{x}m_{y}$, $m_{x}m_{z}$, and $m_{y}m_{z}$ are zero due to their periodic nature. To evaluate these averages we geometrically parametrize~\citep{pinna2013thermally,pinna2014spin} the constant-energy orbit in the anti-parallel well as a function of parameter $\omega$ as
\begin{subequations}
    \begin{align}
        \begin{split}
        &\cosh^{2}{(\omega)} - \sinh^{2}{(\omega)} = 1, \\
        \end{split} \\
        \begin{split}
        &\frac{1}{-g_{L}} m_{x}^{2} - \frac{R}{-g_{L}} m_{z}^{2} = 1,
        \end{split}
\end{align}
\end{subequations}
which leads to the form
\begin{subequations}
    \begin{align}
        \begin{split}
            m_{x}^{c} &= - \sqrt{-g_{L}} \cosh{(\omega)},
        \end{split} \\
        \begin{split}
            m_{z}^{c} &= \pm \sqrt{\frac{-g_{L}}{R}} \sinh{(\omega)},
        \end{split}\\
        \begin{split}
            m_{y}^{c} &= \pm \sqrt{\frac{R-g_{L}}{R}}\sqrt{1 - \zeta^{2}\cosh^{2}{(\omega)}}, 
        \end{split}\\
        \begin{split}
        \zeta^{2} &= -\frac{g_{L}\left(R + 1\right)}{R - g_{L}} = 1 - k^{2},
        \end{split}
    \end{align}    
\end{subequations}
where \textcolor{black}{$|\omega| < \cosh^{-1}{(1/\zeta)}$. Defining time average of some function $p\left(t\right)$ as}
\begin{align}
    \begin{split}
        \langle p\left(t\right) \rangle 
        &=  \frac{4}{T\left(g_{L}\right)} \int_{0}^{t = m_{y}: \sqrt{1+g_{L}} \rightarrow 0} p\left(t\right)\  dt, \\
    \end{split}    
\end{align}
the time averages of $\langle m_{x} \rangle$ and $\langle m_{x}^{2} \rangle$  is evaluated as
\begin{align}
    \begin{split}
        \langle m_{x} \rangle &= \frac{4}{T\left(g_{L}\right)} \int\limits_{0}^{\cosh^{-1}{(1/\zeta)}} m_{x}\left(\omega\right) \Bigg|\frac{\sqrt{R} m_{z}}{R m_{y} m_{z}}\Bigg|\  d \omega \\
        &= -\frac{4}{T\left(g_{L}\right)} \frac{\sqrt{-g_{L}}}{\sqrt{R-g_{L}}}\int\limits_{\mathclap{0}}^{\mathclap{\cosh^{-1}{(1/\zeta)}}}\frac{\cosh{(\omega)} \ d \omega}{\sqrt{1 - \zeta^{2}\cosh^{2}{(\omega)}}} \\
        &= -\frac{4}{T\left(g_{L}\right)} \frac{\sqrt{-g_{L}}}{\zeta\sqrt{R-g_{L}}}\int\limits_{\mathclap{0}}^{\mathclap{\cosh^{-1}{(1/\zeta)}}}  \frac{\cosh{(\omega)} \ d \omega}{\sqrt{\frac{1 - \zeta^{2}}{\zeta^{2}} - \sinh^{2}{(\omega)}}} \\
        &= -\frac{\pi}{2}\frac{{\sqrt{R + -g_{L}}}}{\sqrt{1+R}}\frac{1}{K\left(\zeta\right)} = - \frac{\pi}{2} \sqrt{\frac{R - g_{L}}{R+1}} \frac{1}{K\left(R, g_{L}\right)},
    \end{split}
\end{align}
and 
\begin{align}
    \begin{split}
        \langle m_{x}^{2} \rangle &= \frac{4}{T\left(g_{L}\right)} \int\limits_{0}^{\cosh^{-1}{(1/\zeta)}} m_{x}^{2}\left(\omega\right) \Bigg|\frac{m_{x} \sqrt{R} m_{z}}{m_{x} R m_{y} m_{z}}\Bigg|\  d \omega \\
        &= \frac{-4}{T\left(g_{L}\right)} \frac{g_{L}}{\sqrt{R-g_{L}}}\int\limits_{0}^{\cosh^{-1}{(1/\zeta)}}  \frac{\cosh^{2}{(\omega)} \ dw}{\sqrt{1 - \zeta^{2}\cosh^{2}{(\omega)}}} \\
        &= \frac{-1}{K\left(\zeta\right)} \frac{g_{L}}{\sqrt{1-\zeta^{2}}}\int\limits_{0}^{\cosh^{-1}{(1/\zeta)}}  \frac{\cosh^{2}{(\omega)} \ d \omega}{\sqrt{1 - \frac{\zeta^{2}}{1 - \zeta^{2}}\sinh^{2}{(\omega)}}}\\
        &= \frac{-1}{K\left(\zeta\right)} \frac{g_{L}}{\zeta}\int\limits_{0}^{1}  \frac{\sqrt{1 - \left(1-\frac{1}{\zeta^{2}}\right)u^{2}}\ du}{\sqrt{1 - u^{2}}} \\
        &= \frac{-1}{K\left(\zeta\right)} \frac{g_{L}}{\zeta^{2}} E\left(1-\zeta^{2}\right) = \frac{R - g_{L}}{R+1} \frac{E\left(R, g_{L}\right)}{K\left(R, g_{L}\right)},
    \end{split}
\end{align}
where we have used $dt = \frac{d p\left(\omega\right)/d \omega}{d p\left(t\right)/dt} d \omega$.
Substituting $\langle m_{x} \rangle$ and $\langle m_{x}^{2} \rangle$ in Eq.~(\ref{gdot4}) results in the average rate of energy flow as
\begin{align}\label{gldot}
    \begin{split}
        &\left \langle \frac{\partial g_{L}}{\partial \tau} \right \rangle = -\frac{\pi \alpha}{K\left(R, g_{L}\right)} \sqrt{\frac{R-g_{L}}{1+R}} \left[- \frac{I_{s}}{\alpha} \cos{\phi} \left(1+g_{L}\right) \right. \\
        &+\left. \frac{2}{\pi} \sqrt{1+R}\sqrt{R-g_{L}}\left( E\left(R, g_{L}\right) + g_{L}K\left(R, g_{L}\right) \right)\right]\\
        &- 2\sqrt{\frac{\alpha}{\Delta_{0}}} \sqrt{\frac{\left(R-g_{L}\right)}{K\left(R, g_{L}\right)}} \sqrt{E\left(R, g_{L}\right) + g_{L}K\left(R, g_{L}\right)}\circ \dot{W}_{g_{L}}.
    \end{split}
\end{align}
In this paper, we have considered the spin polarization $\mathbf{\hat{n}}_{p}$ to be collinear to the easy axis $\mathbf{\hat{n}}_{e}$ so $\phi = 0$ which leads to Eq.~(\ref{avg}).

\section{Solving the integral in Eq.~(\ref{tau1})}\label{tau_s}
In Eq.~(\ref{tau1}), substitute $R-g_{L} = u^{2}$, so that we have 
\begin{align}
    \begin{split}
        \tau_{s} &= \frac{1}{2A\alpha} \frac{R+4}{R+2}\int\limits_{\sqrt{R-g_{L_{i}}}}^{\sqrt{R-g_{L_{f}}}} \frac{u^{2} - \frac{R(1+R)}{R+4}}{u^{2} - \frac{R(1+R)}{R+2}}\left[\frac{du }{P(u)}\right],
    \end{split}
\end{align}
where $P(u) = u^{5} - (B/A + 2R)u^{3} + \frac{\tilde{I}_{s}}{A\sqrt{1+R}}u^{2} + (R^{2}+(B/A)R + C/A)u - \tilde{I}_{s}\sqrt{1+R}/A$ and $\lambda_{i}$ are the roots of the polynomial $P(u)$ which are evaluated numerically. Using partial fractions to resolve the denominator of the integral we finally have

\begin{align}
    \begin{split}
    \tau_{s} &= \frac{1}{2\alpha\left(R+2\right)}\vast[\sum_{i = 1}^{5} \frac{N}{D} \log{\left[\frac{(R-g_{L_{f}}) - \lambda_{i}}{(R-g_{L_{i}}) - \lambda_{i}}\right]}\vast. \\
    &+ \left. \sqrt{\frac{R(1+R)}{R+2}} \left\{\frac{\log{\left[\frac{\sqrt{(R-g_{L_{f}})(R+2)} -\sqrt{R(1+R)}}{\sqrt{(R-g_{L_{i}})(R+2)} -\sqrt{R(1+R)}}\right]}}{A\prod\limits_{n=1}^{5} \left(\frac{R(1+R)}{R+2} - \lambda_{i}\right)} \right. \right. \\ 
    &+ \Vast. \left. \frac{\log{\left[\frac{\sqrt{(R-g_{L_{f}})(R+2)} +\sqrt{R(1+R)}}{\sqrt{(R-g_{L_{i}})(R+2)} +\sqrt{R(1+R)}}\right]}}{A\prod\limits_{n=1}^{5} \left(\frac{R(1+R)}{R+2} + \lambda_{i}\right)} \right\} \Vast].
    \end{split}
\end{align}

\vspace{-10pt}
\section{\texorpdfstring{$R \to \infty$}{Lg}} \label{tau_s2}
\vspace{-10pt}
For large values of $R$, Eq.~(\ref{avg}) in the deterministic regime along with rational approximations for the elliptic integrals, \textcolor{black}{Eqs.~(\ref{K}) and~(\ref{E}) can be simplified to}
\begin{align}
    \begin{split}
            \frac{\partial g_{L}}{\partial \tau} &= 4\alpha \left[\frac{1 - g_{L}}{3 - g_{L}}\right] \left[\tilde{I}_{s} \left(1+g_{L}\right) - R\left\{g_{L}\frac{3 - g_{L}}{2 - 2g_{L}} \right. \right.\\
            & \left. \left. + 1 - \frac{(1+g_{L})}{4} \frac{(1+g_{L})^{2} - 28(1+g_{L}) + 64}{4(1+g_{L})^{2} - 40(1+g_{L}) + 64} \right\}\right].
    \end{split}
\end{align}
Now substituting $1+g_{L} = x$ in the previous equation leads us to
\begin{align}
    \begin{split}
          \frac{d x}{d \tau} &= 4\alpha \left[\frac{x - 2}{x - 4}\right] \left[\tilde{I}_{s} x - R\left\{\left(x-1\right)\frac{x - 4}{2x - 4} \right. \right.\\
            & \left. \left. + 1 - \frac{x}{4} \frac{x^{2} - 28x + 64}{4x^{2} - 40x + 64} \right\}\right],
    \end{split}
\end{align}
which can then be simplified and rearranged to an integral of the form
\begin{align}
    \begin{split}
           \int\limits_{1+g_{L_{i}}}^{1+g_{L_{f}}}  \frac{(x-4)(x-8) dx}{x\left[x^{2} - Ex + F\right]} &= \frac{\alpha}{4}\left(16\tilde{I}_{s} - 7R\right)\tau_{s},
    \end{split}
\end{align}
where $E = \frac{160\tilde{I}_{s} - 60R}{16\tilde{I}_{s} - 7R}$ and $F = \frac{256\tilde{I}_{s} -  128R}{16\tilde{I}_{s} - 7R}$. If the roots of the quadratic equation $x^{2} - Ex + F = 0$ are a and b then the switching time $\tau_{s}$ can be evaluated as
\begin{align}
    \begin{split}
        \tau_{s} &= \frac{1}{2\alpha\left[\tilde{I}_{s} - 0.5R\right]}\left[\log{\left[\frac{1+g_{L_{f}}}{1+g_{L_{i}}}\right]}\right. \\
        &+ \left. \frac{b(a-4)(a-8)}{32(a-b)}\log{\left[\frac{1+g_{L_{f}}-a}{1+g_{L_{i}}-a}\right]} \right.  \\
        & -\left.  \frac{a(b-4)(b-8)}{32(a-b)}\log{\left[\frac{1+g_{L_{f}}-b}{1+g_{L_{i}}-b}\right]}\right].
    \end{split}
\end{align}

\vspace{-5pt}
\section{Uniaxial Limit}\label{uni}
\vspace{-10pt}
For the case of uniaxial limit Eq.~(\ref{avg}) in the deterministic regime reduces to
\begin{align}
    \begin{split}
        \left \langle \frac{\partial g_{L}}{\partial \tau} \right \rangle &= 2 \alpha \sqrt{-g_{L}} \left(1 + g_{L} \right) \left(\tilde{I}_{s} - \sqrt{-g_{L}}\right),
    \end{split}
\end{align}
which can then be integrated as
\begin{equation}
    \int\limits_{g_{L_{i}}}^{g_{L_{f}}}\frac{d g_{L}}{\sqrt{-g_{L}}\left(1 + g_{L}\right) \left(\tilde{I}_{s} - \sqrt{-g_{L}}\right)} = 2\alpha \tau_{s}.
\end{equation}
Substituting $g_{L} = -u^{2}$, and using partial fractions to separate the terms in the denominator and integrating with proper limits leads us to
\begin{equation}
    \begin{split}
        \tau_{s} &= \frac{1}{2\alpha\left(\tilde{I}_{s}^{2} - 1\right)}\left[\tilde{I}_{s}\left\{\log{\left[\frac{1+\sqrt{-g_{L_{i}}}}{1+\sqrt{-g_{L_{f}}}}\right]} \right. \right.\\
        &\left. \left. - \log{\left[\frac{1-\sqrt{-g_{L_{i}}}}{1-\sqrt{-g_{L_{f}}}}\right]}\right\} - \log{\left[\frac{1 + g_{L_{i}}}{1 + g_{L_{f}}}\right]} \right. \\
        &+ \left. 2\log{\left[\frac{\frac{\tilde{I}_{s}}{\alpha} - \sqrt{-g_{L_{i}}}}{\frac{\tilde{I}_{s}}{\alpha} - \sqrt{-g_{L_{f}}}}\right]}\right].
    \end{split}
\end{equation}

\vspace{-15pt}
\section{Eqs.~(\ref{equib_energy})
and~(\ref{cdf})}\label{rho_gL}
\vspace{-10pt}
\textcolor{black}{If a random variable $X$ has certain probability $P_{X}[x]$ then the probability for a random variable $Y = -X^{2}$ could be obtained as}
\begin{flalign}
    \begin{split}
        P_{Y}\left[y\right] &= \mathbb{P}\left[Y \leq y\right] = \mathbb{P}\left[ -X^{2} \leq y\right] \\
        & = 1 - \mathbb{P}\left[ |X| \leq \sqrt{-y}\right] \\ 
        & = 1 - \mathbb{P}\left[-\sqrt{-y} \leq X \leq \sqrt{-y}\right]\\
    \end{split}
\end{flalign}
\textcolor{black}{Now, if we consider that $X$ represents the distribution of magnetization $m_{x}$ while $Y$ represents that of energy $g_{L}$ then for $R \to 0$ using Eq.~(\ref{equib}) one can obtain the probability, }
\begin{flalign}\label{cdf_app}
    \begin{split}
        &P[g_{L}] = 1 - \frac{2}{Z(\Delta_{0}, 0)} \int\limits_{0}^{\sqrt{-g_{L}}} d x \ \exp\left(\Delta_{0} m_{x}^{2}\right) \\
        & = 1 - \frac{2}{Z(\Delta_{0}, 0)\sqrt{\Delta_{0}}} \int\limits_{0}^{\sqrt{-g_{L}}} d m_{x} \ \sqrt{\Delta_{0}} \ \exp\left(\Delta_{0} m_{x}^{2}\right) \\
        & = 1 - \frac{2\exp(\Delta_{0}g_{L})}{Z(\Delta_{0}, 0)\sqrt{\Delta_{0}}\exp(\Delta_{0}g_{L})} \int\limits_{0}^{\sqrt{-\Delta_{0}g_{L}}} du \ \exp\left(u^{2}\right) \\
        & = 1 - \frac{2\exp(-\Delta_{0}g_{L})}{Z(\Delta_{0}, 0)\sqrt{\Delta_{0}}}F\left(\sqrt{-\Delta_{0}g_{L}}\right).
    \end{split}
\end{flalign}
Here we have substituted $\sqrt{\Delta_{0}} m_{x} = u$, and used $F(x) = \exp(-x^{2})\int_{0}^{x} dy \exp(y^{2})$, the Dawson's integral. Since $P[-1] = 0$, therefore, 
\begin{equation}\label{partition2}
    Z(\Delta_{0}, 0) = \frac{2\exp(\Delta_{0})}{\sqrt{\Delta_{0}}}F\left(\sqrt{\Delta_{0}}\right).
\end{equation}
Eqs.~(\ref{cdf_app}) and~(\ref{partition2}) together lead to Eq.~(\ref{cdf}).

In order to arrive at the probability distribution function (PDF) we now differentiate Eq.~(\ref{cdf_app}) with respect to $g_{L}$ to get
\begin{flalign}
    \begin{split}
        \rho_{Y}[g_{L}] & = \frac{1}{Z(\Delta_{0}, 0)} \frac{\exp\left(-\Delta_{0} g_{L}\right)}{\sqrt{-g_{L}}} \\
        & = \frac{\sqrt{\Delta_{0}}}{2 F\left(\sqrt{\Delta_{0}}\right)} \frac{\exp\left(-\Delta_{0} \left(1+g_{L}\right)\right)}{\sqrt{-g_{L}}},
    \end{split}
\end{flalign}
where we have used $Z(\Delta_{0}, R) = Z(\Delta_{0}, 0)$.

\nocite{*}
\bibliography{aipsamp}

\end{document}